\documentclass[preprint2]{aastex}
\usepackage{amsmath}

\slugcomment{submitted to ApJ main journal}
\shorttitle{X-ray Emission Properties of Large Scale Jets}
\shortauthors{Kataoka \& Stawarz}

\begin{document}

\title{
X-ray Emission Properties of Large Scale Jets, Hotspots and
Lobes in Active Galactic Nuclei
}

\author{Jun Kataoka$^{1}$ and \L ukasz Stawarz$^{2}$}
\affil{$^1$Tokyo Institute of Technology, Meguro, Tokyo, Japan}
\affil{$^2$Obserwatorium Astronomiczne, Uniwersytet Jagiello\'nski, Krak\'ow, Poland}

\altaffiltext{1}{e-mail:kataoka@hp.phys.titech.ac.jp}
\altaffiltext{2}{e-mail:stawarz@oa.uj.edu.pl}

\begin{abstract}

We examine a systematic comparison of jet-knots, hotspots and radio lobes recently observed with $Chandra$ and $ASCA$. This report will discuss the origin of their X-ray emissions and investigate the dynamics of the jets. The data was compiled at well sampled radio (5 GHz) and X-ray frequencies (1keV) for more than 40 radio galaxies. We examined three models for the X-ray production: synchrotron (SYN), synchrotron self-Compton (SSC) and external Compton on CMB photons (EC). For the SYN sources --- mostly jet-knots in nearby low-luminosity radio galaxies --- X-ray photons are produced by ultrarelativistic electrons with energies 10$-$100 TeV that must be accelerated in situ. For the other objects, conservatively classified as SSC or EC sources, a simple formulation of calculating the ``expected'' X-ray fluxes under an equipartition hypothesis is presented. We confirmed that the observed X-ray fluxes are close to the expected ones for non-relativistic emitting plasma velocities in the case of radio lobes and majority of hotspots, whereas considerable fraction of jet-knots is too bright at X-rays to be explained in this way. We examined two possibilities to account for the discrepancy in a framework of the inverse-Compton model: (1) magnetic field is much smaller than the equipartition value, and (2) the jets are highly relativistic on kpc/Mpc scales. We concluded, that if the inverse-Compton model is the case, the X-ray bright jet-knots are most likely far from the minimum-power condition. We also briefly discuss the other possibility, namely that the observed X-ray emission from all of the jet-knots is synchrotron in origin.

\end{abstract}

\keywords{galaxies: jets --- magnetic fields --- radiation mechanism: non-thermal}

\section{Introduction}

The excellent spatial resolution of $Chandra$ X-ray Observatory has opened a new era to study the large scale jets in powerful extragalactic radio sources. At the time of this writing, more than 40 radio-loud AGNs are known to possess X-ray counterparts of radio jets on kpc to Mpc scales (Harris \& Krawczynski 2002, Stawarz 2004 and references therein; see also \texttt{http://hea-www.harvard.edu/XJET/}). Bright X-ray knots (hereafter ``jet-knots'') are most often detected, but the X-ray emissions from the hotspots and radio lobes are also reported in a number of FR II radio galaxies and quasars (e.g., Wilson, Young \& Shopbell 2000; 2001; Hardcastle et al. 2002b; 2004, Tashiro et al. 1998; Isobe 2002).

The broad-band spectra of jet-knots, hotspots, and lobes detected by $Chandra$ show great variety between radio and X-ray energy bands. In nearby FR I sources, typical X-ray-optical-radio spectrum of the jet-knots is consistent with a single smoothly broken power-law continuum, suggesting that this broad-band emission is entirely due to non-thermal synchrotron radiation from a single electron population (e.g., Marshall et al. 2002 and Wilson \& Yang 2002 for M~87). In most other sources, however, the X-ray knots' spectra are much harder than expected from a simple extrapolation of the radio-to-optical fluxes. These situations are believed that both the radio and optical emissions are due to synchrotron radiation, whereas X-ray photons are produced via the inverse-Compton scattering of either synchrotron photons (SSC) or cosmic microwave background photons (EC; Tavecchio et al. 2000, Celotti, Ghisellini \& Chiaberge 2001). Other (synchrotron) models have been also proposed to explain intense X-ray emission of the large-scale quasar jets (e.g., Dermer \& Atoyan 2002, Stawarz \& Ostrowski 2002). In the case of the hotspots in powerful sources one finds an analogous controversy regarding the X-ray emission: although in many objects this emission is consistent with the standard SSC model (see, e.g., Wilson et al. 2000 for Cygnus~A), in some other sources it cannot be simply explained in this way, suggesting most likely a synchrotron origin of the detected X-ray photons (see, e.g., Hardcastle et al. 2004). For the extended lobes of quasars and FR IIs the X-ray radiation is established to be produced by the EC process involving CMB target radiation. In some cases, however, infrared target photons from quasar cores may contribute to the inverse-Compton lobes' emission at keV photon energy range (Brunetti, Setti \& Comastri 1997).

In the standard picture of FR II radio galaxies and quasars, the
relativistic jet is decelerated in a hotspot converting part of its
energy into relativistic electrons and part in magnetic field. Then the
shocked plasma moves inside the head region just behind the hotspot, and
expands almost adiabatically to form diffuse, extended radio lobes. Even
though this picture appears to be simple, much of the fundamental
physics behind it remains unclear (see, e.g., recent monograph by De
Young 2002a). For example, the velocity and dynamics of the large-scale
jets is unknown. From the analogy to sub-pc jets in blazar-type AGNs, it
is plausible that some of the FR II and quasar jets are highly
relativistic even on kpc/Mpc scales. Recent studies on the optical
emission of the large-scale jets seem to justify this hypothesis (e.g.,
Sparks et al. 1995, Scarpa \& Urry 2002, Jester 2003), and the usually
discussed versions of the EC model for the X-ray jet-knots indeed
require the jet bulk Lorentz factors $\Gamma_{\rm BLK} \geq 10$ (e.g.,
Harris \& Krawczynski 2002). Yet, the exact velocity structure both
along and across large-scale jets in FR II radio galaxies and quasars
remains an open issue. The strong terminal shocks at the hotspots are
unlikely to be moving with high bulk Lorentz factors, but moderately
relativistic motions ($\Gamma_{\rm BLK}$ $\le$ a few) are permitted by
hydrodynamic simulations (e.g., Aloy et al. 1999). We note, that such
simulations repeatedly reveal a complex hotspots' morphology, especially
at the late stages of the jet evolution (e.g., Marti et al. 1997,
Mizuta, Yamada \& Takabe 2004). Finally, the  main-axis expansion of
radio lobe is thought to be sub-relativistic; 
$\Gamma_{\rm BLK}$ $\simeq$ 1. However, detailed transport and spatial 
distribution of the radiating particles within the lobes of powerful
radio sources is still being debated (e.g., Blundell \& Rawlings 2000, 
Kaiser 2000, Manolakou \& Kirk 2003).

As for the velocity of jet plasma, the strength of magnetic field in radio galaxies is an open matter. Assuming an equipartition field value in the lobes (1$-$10 $\mu$G), which seems to be supported by the X-ray lobes' observations, a simple flux conservation argument predicts the magnetic field in the jets as high as 0.01$-$1 G (De Young 2002b). Such a strong magnetic field is problematic, since numerical simulations of Poynting-flux dominated jets (e.g., Komissarov 1999) cannot correctly reproduce the observed large-scale morphologies of powerful radio sources. Thus, an amplification of the magnetic field to the equipartition value in strong jet terminal shock and in its turbulent downstream region is required, although only little theoretical investigations of this issue has been reported (see De Young 2002b). Let us mention in this context, that turbulent processes that may lead to amplification of the magnetic field can manifest in formation of the flat-spectrum synchrotron X-ray features, such like the ones discovered recently in the hotspots of Cygnus A radio galaxy (Ba\l uci\'nska-Church et al. 2004). On the other hand, the equipartition of energy between the magnetic field and the radiating electrons, established for some high-luminosity sources, may not be valid in general, especially in the case of low-luminosity hotspots (Hardcastle et al. 2004). Finally, we note that the configuration of the magnetic field within the lobes is also not well understood (see a discussion in Blundell \& Rawlings 2000).

Unfortunately, present radio-to-X-ray observations are not sufficient to discriminate conclusively between different models proposed in order to explain multiwavelength emission of the large-scale structures of powerful radio sources, and of their kpc/Mpc jets in particular. However, we believe that a systematic comparison between broad-band radiative properties of the jet-knots, hotspots, and lobes will provide important clues to dynamics and the physics of large scale jets, and to put some constraints on the models discussed in the literature. Keeping these motivations in mind, the purpose of this paper is to obtain a rough, but unified picture which may link the jet-knots, hotspots and radio lobes, rather than modeling individual sources in a sufficiently detailed manner. Obviously, detailed studies on individual cases are irreplaceable. In fact, many controversial issues briefly touched in this analysis will remain open until such detailed investigations, based on long multiwavelength observations, are performed. We emphasize, that our analysis confirms many results known from the literature (see, e.g., Stawarz 2004 for a review), although for a large number of sources modeled in addition in a uniform way. Basing on this homogeneous approach, we explore however some new, hardly discuss in the literature aspects of the physics behind the X-ray emission models for the considered objects. Let us also mention, that in this paper we do not consider hadronic models for the broad-band emission of the large-scale jets and their hotspots (see, e.g., Aharonian 2002, Atoyan \& Dermer 2004).

Our presented study is based on data analysis for a sample consisting of  26 radio galaxies, 14 quasars, and 4 blazars. We collected all existing data at well sampled  radio (5 GHz) and X-ray (1 keV) frequencies and analyzed them in a systematic manner. In $\S$2, we defined sample selection and observables used in this paper. In $\S$3, we presented a simple formulation of calculating the ``expected'' X-ray flux densities for the SSC and EC models taking the relativistic beaming effect into account. We then compared the physical quantities (beaming factor and the magnetic field) of the jet-knots, hotspots, and lobes. In $\S$4 we discuss the results and the summary is presented in $\S$5.

\section{Data and analysis}

\subsection{Sample}

Table 1 compiles a list of ``X-ray jet sources'' in which jet-knots,
hotspots and/or radio lobes are detected by $Chandra$ and $ASCA$. The
first pioneer work have been reported by Harris \& Krawczynski (2002),
where the emission mechanisms of 18 X-ray jet sources (mainly jet-knots)
are discussed in the framework of relativistically moving jet
model. They continue to maintain current information at
\texttt{http://hea-www.harvard.edu\\
/XJET/}, which conveniently summarize
the name, coordinate, distance, and morphology of the X-ray jet
sources. Our sample contains all of the sources listed in this page,
with additional information on the X-ray observations of radio lobes
mainly organized by $ASCA$.

Before compiling the data, we have performed quick re-analysis of $Chandra$ data (if already archived) to check the published results, and found no discrepancy. We therefore refer to published results (fluxes and spectral indices) unless otherwise stated in this paper. This gives a large number of objects known to us as of 2004 June, which contains 44 X-ray jet sources (56 jet-knots, 24 hotspots, and 18 radio lobes: see Table 2). We are aware that  our sample is still incomplete, as the known X-ray jet sources are increasing their number day by day. Nevertheless such a list provides a convenient overview of X-ray jet sources detected so far, and provide a useful hint to $predict$ fluxes of unobserved X-ray jet sources. We also note that Hardcastle et al. (2004) recently summarizes the X-ray emission properties of the hotspots in FR II radio galaxies.
 
The  basic information about each source are listed in Table 1: (1) source name, (2) redshift $z$, (3) luminosity distance to the source $d_{\rm L}$ adopting  $H_0$ = 75 km s$^{-1}$ Mpc$^{-1}$ and $q_0$ = 0.5, (4) classification, and (5) references. RG denotes radio galaxy of either Fanaroff \& Riley type I (FR I) or type II (FR II), QSO denotes either core dominated (CD) quasars or lobe dominated (LD) quasars, and BLZR denotes blazar-class.

More detailed information on each source are listed  in Table 2. In the second column we denote ``knot (K)'' to indicate a distinct structure in the jet, ``hotspots (HS)'' as a terminal bright enhancement at an end of the FR II jet or as one of the multiple features associated with a termination of the jet, and  ``lobe (L)'' as a diffuse extended structure associated with a radio lobe.  A suffix after K, HS, and L means the identification of each structure. For example ``K-A'' denotes `knot-A'' and ``HS-SE'' means ``South-east hotspot''. In succeeding columns of Table 2, we have listed 6 observables: (1) $\alpha_{\rm R}$: radio spectral index measured at 5 GHz, (2) $f_{\rm R}$: radio flux density at 5 GHz in mJy, (3) $\alpha_{\rm X}$: X-ray spectral index at 1 keV, (4) $f_{\rm X}$: X-ray flux density at 1 keV in nJy, (5) $f_{\rm O}$: optical flux density at 5$\times$10$^{14}$ Hz in $\mu$Jy, and (6) $\theta$: radial size of the emitting region in arcsec. When observations have not been reported at 5 GHz or 1 keV, we calculate the flux by extrapolating the nearest measured frequency by assuming the spectral index listed in the table. A suffix $f$ means that we have assumed the fixed value for this calculation.

\subsection{Radio/X-ray comparison}

Figure 1 shows the distribution of the spectral indices in the radio band ($\alpha_{\rm R}$; $upper$) and in the X-ray band ($\alpha_{\rm X}$; $lower$), respectively. Note that radio spectral index shows a relatively narrow distribution centered at 0.8 and there is no clear difference between the jet-knots, hotstpots and radio lobes. As is widely believed, the radio emissions of these sources are most likely due to the synchrotron radiation from the low-energy population of relativistic electrons. In other words, energy index of accelerated electron is narrowly distributed around $s$ = ($\alpha_{\rm R}$ +1)/2 $\simeq$ 2.6, which is slightly steeper than the one expected from a diffusive acceleration at nonrelativistic shocks, $s$ = $2$. Let us note in this context, that analytical and numerical studies of particle acceleration at relativistic shocks (reviewed by, e.g., Kirk \& Duffy 1999 and Ostrowski 2002), indicate that in such a case one can expect variety of particle spectra, with the asymptotic power-law inclination $s$ = $2.2$ for the strong turbulence condition and ultrarelativistic shock velocity. We also note, that stochastic second-order Fermi processes do not favor any universal value of the power-law spectral index characterizing accelerated electrons.

\begin{figure}[htb]
\includegraphics[angle=90,scale=.40]{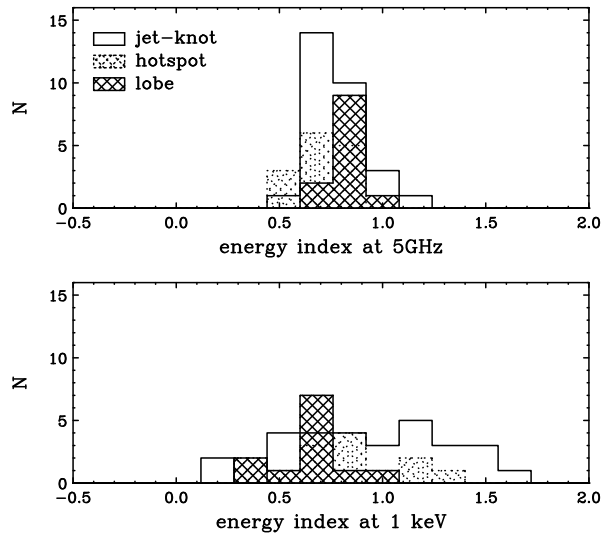}
\caption{Distribution of the energy index measured at 5 GHz and at 1 keV.}
\end{figure}

Meanwhile, the X-ray energy index, $\alpha_{\rm X}$, is widely distributed from 0.2 to 1.6. Part of the reason may be relatively large uncertainties in determining the spectral shape of faint X-ray sources compared to the radio spectral shape, but even if only bright (i.e., small error bars) X-ray sources are plotted, the same trend is obtained. Steep X-ray sources are most frequently found in nearby FR I radio galaxies. As discussed in the literature, the X-ray fluxes in these sources may smoothly connect with radio/optical fluxes and hence are considered to be the highest energy tail of the synchrotron radiation. For the X-ray emission from other jet-knots the situation is less clear. Flat X-ray spectral indices may indicate pile-up effects at the high-energy part of the electron energy distribution, advocating thus synchrotron origin of the keV photons (see Harris, Mossman \& Walker 2004), or, oppositely, spectral flattenings occurring at the low-energy part of the electron continuum thus being consistent with the EC interpretation of the X-ray knots' emission. Clearly, spectral information alone are not sufficient at the moment to distinguish either between synchrotron and inverse-Compton origin of the keV photons from the jet-knots in most of the cases, or to indicate the appropriate particle acceleration process.

\begin{figure}[htb]
\includegraphics[angle=90,scale=.40]{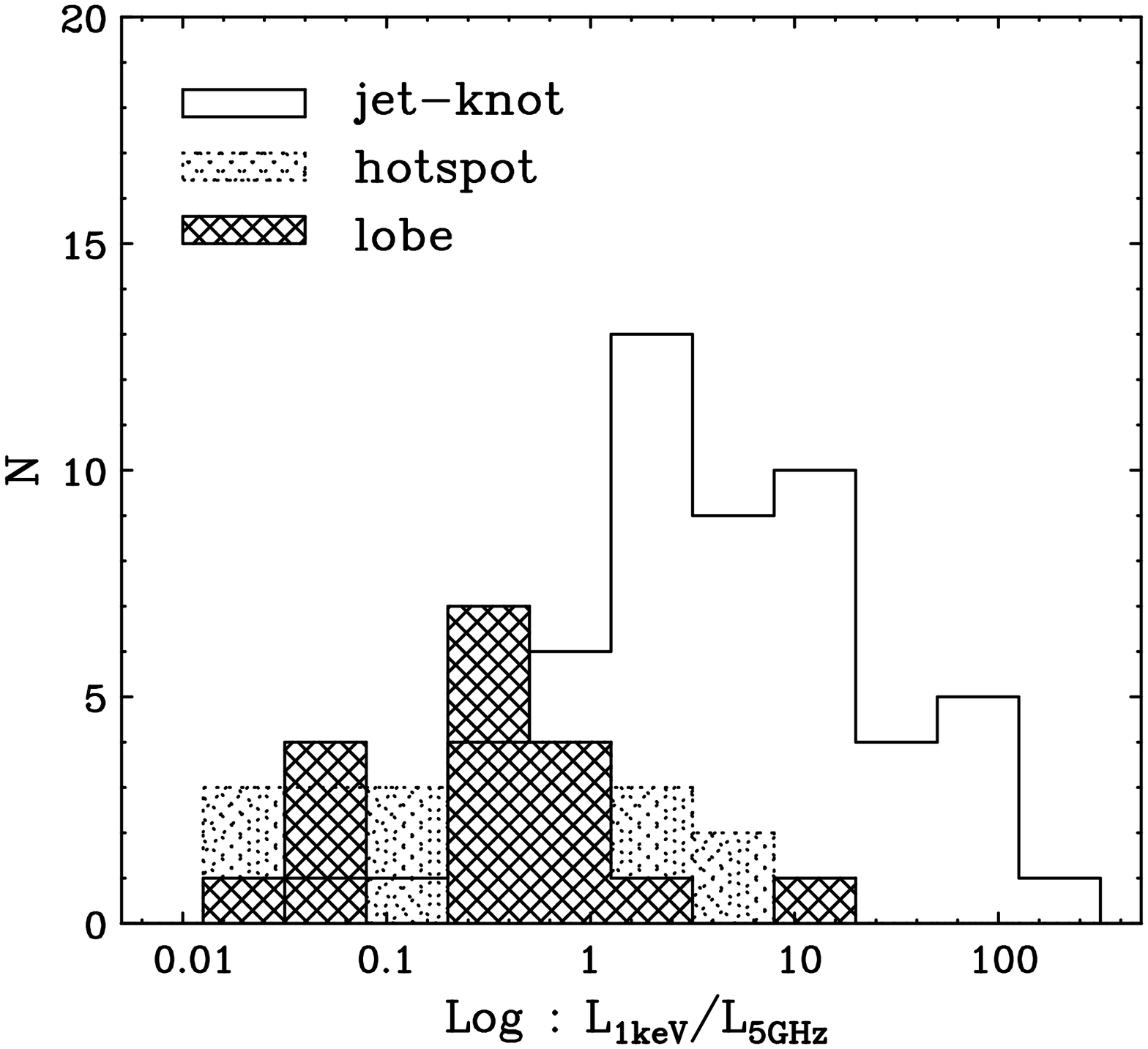}
\caption{Distribution of the ratio between $L_{\rm 1keV}$ and $L_{\rm 5GHz}$.}
\end{figure}

Figure 2 presents the distribution of luminosity ratio of $L_{\rm R}$
and $L_{\rm X}$, where $L_{\rm R}$ = 4$\pi$$d_{\rm L}^2$$f_{\rm
R}$$\nu_{\rm R}$ and  $L_{\rm X}$ = 4$\pi$$d_{\rm L}^2$$f_{\rm
X}$$\nu_{\rm X}$, respectively. Note that a clear difference can be seen between the jet-knot and hotspot or radio lobes. The jet-knots tend to be  much brighter in X-rays compared to the hotspots and radio lobes. This trend is seen more clearly in Figure 3, where the correlation between $L_{\rm R}$ and $L_{\rm X}$ is plotted in two dimensional space. One finds several important tendencies which cannot be accounted by the sampling bias effect. First, hotspots and radio lobes occupy only the high-luminosity part of the plot, namely $\ge$ 10$^{40}$ erg s$^{-1}$. Secondly, low-luminosity hotspots tend to be brighter in X-ray, as has been pointed out by Hardcastle et al. (2004). Thirdly, $L_{\rm R}$ $\ge$ $L_{\rm X}$ for most of the hotspots and radio lobes, but most of the jet-knots show an opposite trend.

\begin{figure}[htb]
\includegraphics[angle=90,scale=.40]{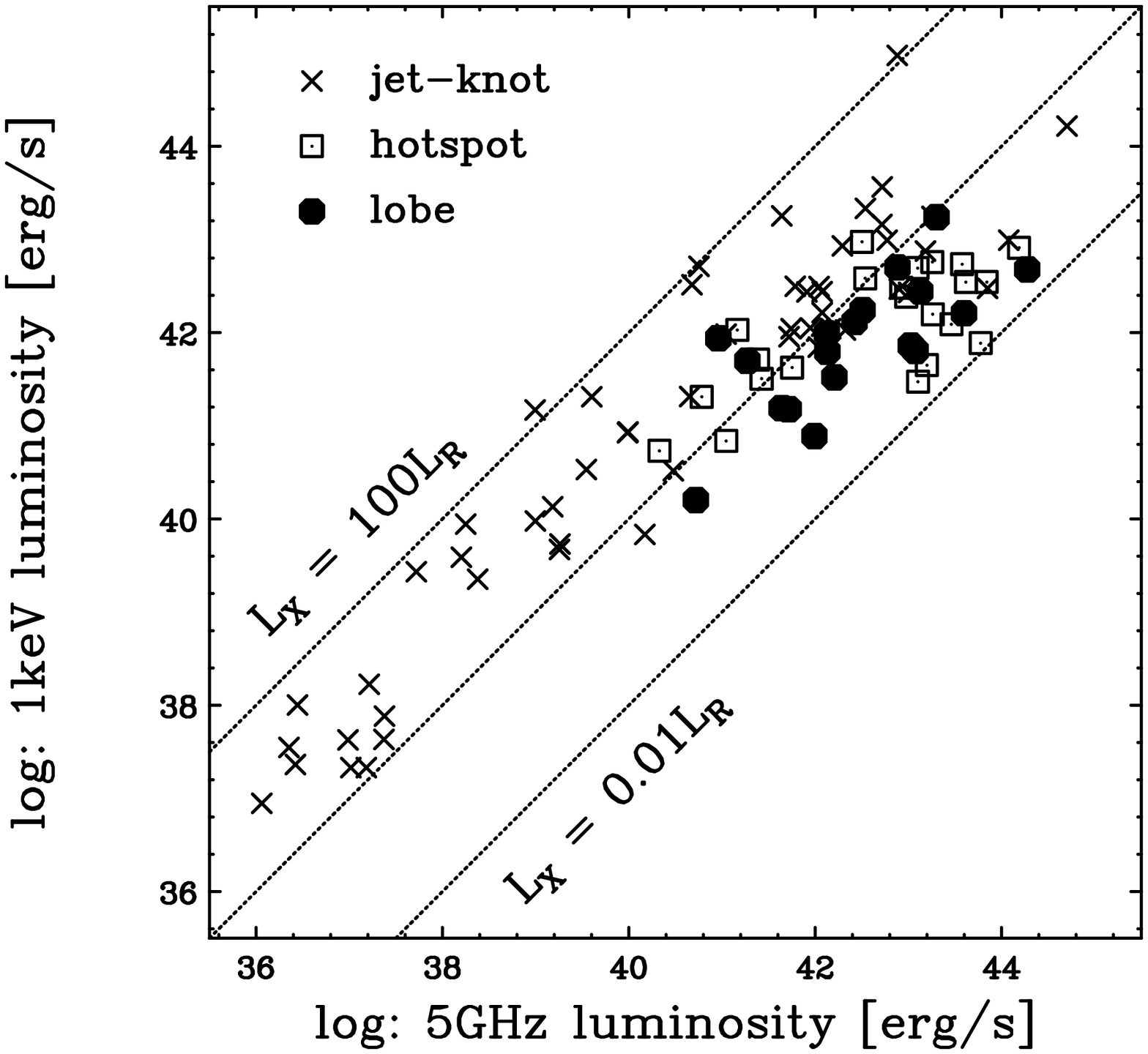}
\caption{Relation between the luminosities $L_{\rm 5GHz}$ and $L_{\rm 1keV}$.}
\end{figure}

We should note that due to limited sensitivity of $Chandra$ (typically 0.1 nJy at 1 keV for 10 ksec exposure), we would not expect to detect the X-ray emission from the ``X-ray faint'' jet-knots. Therefore the lack of the X-ray faint (i.e., $L_{\rm R}$ $\ge$ $L_{\rm X}$) jet-knots at the bottom left corner of Figure 3 would be biased by the sensitivity of $Chandra$ detector. In fact, we can find a few X-ray faint jet-knots at the top right corner, where the luminosity is the highest. However, even if only high-luminosity sources are selected, we can see a clear difference between the jet-knots, hotspots and radio lobes, namely ``X-ray bright'' sources are found only in jet-knots. Apparently, this is not due to the sampling effect since we certainly would have been able to detect ``X-ray bright'' hotspots if they existed.

\section{Model application to data}

In this section we present a simple formulation of computing an equipartition magnetic field strength $B_{\rm eq}$ from an observed radio flux $f_{\rm R}$ measured at a radio frequency $\nu_{\rm R}$. Next, we calculate the ``expected'' SSC and EC luminosities for $B_{\rm eq}$, to compare them with the observed X-ray luminosities. In the analysis, we include possible relativistic bulk velocity of the jet plasma. Taking the obtained results into account, and analyzing additionally the observed broad-band spectral properties of the compiled sources (including optical fluxes), we follow the ``conservative'' classification of the compiled X-ray sources into three groups, namely (i) synchrotron involving single/broken power-law electron energy distribution (SYN), (ii) synchrotron self-Compton (SSC) and (iii) external Compton of CMB photons (EC). Finally, we discuss the validity of the applied classification scheme, and compare it briefly with the classification introduced in the literature.

\vspace{5mm}

\subsection{Equipartition magnetic field}

In order to determine the X-ray emission properties of the large-scale jets, we first estimated the magnetic field strength under the minimum-power hypothesis using the observed radio luminosities measured at 5 GHz. As we have reviewed in $\S$1, it is generally believed that the magnetic field energy density $u_{\rm B}$ and the particle energy density $u_{\rm e}$ may be close to the equipartition in a number of radio sources. Therefore our approach is that we first assume an equipartition to calculate ``predicted'' inverse-Compton X-ray luminosities, and then compare them to the observed ones. If a large discrepancy occurs, this may suggest that equipartition is strongly violated, that the inverse-Compton origin of the observed keV photons is not the case, or that we have to consider another correction factor, such as Doppler beaming factor $\delta$, as we will discuss below.

Since the synchrotron luminosity, $L_{\nu}$, is proportional to $u_{\rm e}$$u_{\rm B}$$V$, where $V$ is the volume of the emitting regions, we can estimate the equipartition magnetic field $B_{\rm eq}$ for a given luminosity observed at a radio frequency $\nu$. Under the assumption of $no$ relativistic beaming ($\delta$ = 1), $B_{\rm eq}$ is expressed as

\begin{equation}
\begin{aligned}
B_{\rm eq, \, \delta=1} = \left[ \frac{3 \mu_0}{2} \frac{G(\alpha) \eta L_\nu}{V}   \right]^{2/7}\\
 \propto \left(  \frac{\eta L_\nu}{V}  \right)^{2/7} \nu^{1/7} \quad , 
\end{aligned}
\end{equation}

\noindent
where $\mu_0$ is permeability of free space, $G(\alpha)$ is a function given  in Longair (1994), $\alpha$ is the spectral energy index and $L_{\nu}$ is the synchrotron luminosity  measured at a frequency $\nu$. $\eta$ is the ratio of energy density carried by proton and electrons to the energy density of the electrons, i.e.,  $\eta$ = 1 for the leptonic ($e^{-}$$e^{+}$) jet and $\eta$ = 1836 for the hadronic ($e^{-}$$p^{+}$) jet for which the ratio of proton to electron energy densities equals the ratio of their rest masses. In the last approximation in the equation (1) we put minimum synchrotron frequency $\nu_{\rm min}$ = $\nu$ and $\alpha$ = 0.75. The latter choice is justified by a narrow distribution of the radio spectral indices in the compiled dataset (Figure 1, $upper$). The former choice gives the $minimum$ value of $B_{\rm eq}$ for the observed $L_{\nu}$ at some given frequency $\nu$. Below we consider $\nu$ = $\nu_{\rm R}$ = 5 GHz, although it is obvious that the minimum radio frequency has to be lower than this (especially in the case of the EC model, which requires presence of low-energy electrons with energies below GeV). However, the difference between equipartition magnetic field computed for $\nu_{\rm min}$ = $\nu$ and for $\nu_{\rm min}$ $\neq$ $\nu$, respectively, is rather small, $\propto$ $(\nu_{\rm min} / \nu)^{1/14}$. In addition, the expected spectral flattenings at the low-energy part of the synchrotron continuum are likely to make this difference even smaller.

In general, an emission volume, $V$, is quite uncertain for astrophysical sources due to the limited angular resolution of detectors and projection effect. We have assumed that the emitting region has a spherical volume of a certain angular radius $\theta$[''] for all the jet structures. This is obviously an over-simplified assumption, however it significantly reduces the complexity of the models. Most of the jet-knots and hotspots are point-like sources when observed with $Chandra$. We therefore set an upper limit of $\theta$ = 0.3'', unless there are additional radio/optical observations obtained with better angular resolution. Meanwhile a number of radio lobes show extended structures, but morphology is more complicated than a homogeneous sphere. We therefore calculated the volume by assuming a cylinder or a rotational ellipse, and then approximated it as a sphere which has an equal volume.

For a relativistically moving plasma, the equipartition magnetic field measured in the emitting plasma rest frame is related to the equipartition value computed for no beaming by the relation (Stawarz, Sikora \& Ostrowski 2003)

\begin{equation}
\begin{aligned}
B_{\rm eq} =  B_{\rm eq, \, \delta =1} \delta^{-5/7}.
\end{aligned}
\end{equation}

\noindent
The above expression can be more conveniently written as

\begin{equation}
\begin{aligned}
B_{\rm eq} = 123 \hspace{3mm} \eta^{2/7} (1+z)^{11/7} \left( \frac{d_{\rm L}}{\rm 100 Mpc} \right)^{-2/7}\\
\times\left( \frac{\nu_{\rm R}}{\rm 5GHz} \right)^{1/7}
\left( \frac{f_{\rm R}}{100{\rm mJy}} \right)^{2/7}
\left( \frac{\theta}{0.3''} \right)^{-6/7}\\ 
\times\delta^{-5/7} {\rm [\mu G]},
\end{aligned}
\end{equation}

\noindent
where $d_{\rm L}$ is the luminosity distance to the source, and $f_{\rm R}$  is the observed radio luminosity measured at frequency $\nu_{\rm R}$. $B_{\rm eq, \, \delta=1}$ for various jet sources are calculated in Table 2 for $\eta = 1$, what gives again the $minimum$ value of $B_{\rm eq, \, \delta = 1}$. We note that this particular choice does not refer exclusively to the leptonic jet model. For example, it may refer to the case of energy equipartition between jet magnetic field and radiating electrons solely. We note, that the discussion on the jet composition is still open, and the situation may be quite complex as, for example, the jet can be composed predominantly from the $e^{-}$$e^{+}$-pairs, but still remain dynamically dominated by the (cold) hadrons (see Sikora \& Madejski 2000).

\vspace{5mm}

\subsection{Synchrotron (SYN) case}

We first consider the case where the X-ray emissions are due to the synchrotron radiation emitted by the electrons with the Lorentz factor $\gamma_{\rm X}$. We assume that the magnetic field in the jet moving plasma is close to equipartition $B_{\rm eq}$, and its relativistic beaming factor is $\delta$. Then the observed X-ray frequency is given by

\begin{equation}
\begin{aligned}
\nu_{\rm X} \simeq 1.2 \times 10^6  \gamma_{\rm X}^2  B_{\rm eq} (1+z)^{-1}
\delta\\
\simeq 1.2 \times 10^6 \gamma_{\rm X}^2  B_{\rm eq, \delta=1} (1+z)^{-1}
 \delta^{2/7}.
\end{aligned}
\end{equation}

\noindent
The respective electron Lorentz factor, $\gamma_{\rm X}$, is hence given as

\begin{equation}
\begin{aligned}
\gamma_{\rm X} \simeq 4.5 \times 10^7  \left( \frac{\nu_{\rm X}}{\nu_{\rm 1keV}}  \right)^{1/2} \left( \frac{B_{\rm eq, \delta=1}}{100 {\rm \mu G}} \right)^{-1/2}\\
\times(1+z)^{1/2} \delta^{-1/7},
\end{aligned}
\end{equation}

\noindent
where ${\nu_{\rm 1keV}}$ is 2.4$\times$10$^{17}$ Hz. Therefore, even though $\delta$ is quite uncertain, the estimated value of $\gamma_{\rm X}$ is not affected significantly since $\gamma$ is only weakly dependent on $\delta$, i.e., $\propto$ $\delta^{-1/7}$.

\vspace{5mm}

\subsection{Synchrotron self-Compton (SSC) emission}

The observed radio frequency is approximately

\begin{equation}
\begin{aligned}
\nu_{\rm R} \simeq 1.2 \times 10^6 B_{\rm eq} \gamma_{\rm R}^2  (1+z)^{-1} \delta ,\\
\end{aligned}
\end{equation}

\noindent
where $\gamma_{\rm R}$ is a Lorentz factor of electrons which emit synchrotron photons at $\nu_{\rm R}$. In the SSC case, electrons upscatter synchrotron photons to a frequency

\begin{equation}
\begin{aligned}
\nu_{\rm IC} \simeq \frac{4}{3}  \gamma_{\rm R}^2 \nu_{\rm R} = 2.8 \times10^{17} \left( \frac{\nu_{\rm R}}{\rm {5 GHz}}  \right)^2\left(  \frac{B_{\rm eq}}{100{\rm \mu G}}  \right)^{-1}\\
\times(1+z)  \delta^{-1} \quad {\rm [Hz]} \\ 
= 2.3 \times 10^{17} \eta^{-2/7} (1+z)^{-4/7} \left( \frac{d_{\rm L}}{\rm 100 Mpc} \right)^{2/7}\\
\times\left(\frac{\nu_{\rm R}}{\rm 5GHz} \right)^{13/7}
\left( \frac{f_{\rm R}}{100 {\rm mJy}} \right)^{-2/7}\left( \frac{\theta}{0.3''} \right)^{6/7}\\
\times\delta^{-2/7} \quad {\rm [Hz]}.
\end{aligned}
\end{equation}

\noindent
Note that, $\nu_{\rm IC}$ depends both on the observed radio frequency
and magnetic field strength $B_{\rm eq}$. To calculate the X-ray flux at
an observed frequency $\nu_{\rm X}$, we have to extrapolate the
inverse-Compton flux calculated for $\nu_{\rm IC}$ by assuming the
observed X-ray spectral index $\alpha_{\rm X}$. In the SSC case,  we
expect $\alpha_{\rm X}$ $\simeq$ $\alpha_{\rm R}$, if the synchrotron
continuum can be well approximated by a single power-law with $\alpha$
$\simeq$ $\alpha_{\rm R}$. The ratio of X-ray (SSC) luminosity to the 
radio (synchrotron) luminosity is therefore

\begin{equation}
\begin{aligned}
\frac{\nu_{\rm IC} f_{\rm IC}}{\nu_{\rm R} f_{\rm R}}   \simeq  \frac{\nu_{\rm X} f_{\rm X}}{\nu_{\rm R} f_{\rm R}} \left( \frac{\nu_{\rm IC}}{\nu_{\rm X}} \right)^{1 - \alpha_{\rm R} } \simeq   \frac{u'_{\rm sync}}{u'_B}, 
\end{aligned}
\end{equation}

\noindent
where $u'_{\rm sync}$ and $u'_B$ are  the synchrotron photon energy density and the magnetic field density, respectively, both evaluated in the emitting region rest frame denoted by primes. $u'_{\rm sync}$ is given by

\begin{equation}
\begin{aligned}
u'_{\rm sync} = \frac{d_{\rm L}^2 \nu_{\rm R} f_{\rm R}}{R^2 c \delta^4} = 7.9 \times 10^{-13} (1+z)^4 \left( \frac{\nu_{\rm R}}{5 {\rm GHz}}  \right)\\
\times \left( \frac{f_{\rm R}}{100 {\rm mJy}} \right) \left(\frac{\theta}{0.3''}  \right)^{-2} \delta^{-4} \quad \rm{[erg/cm^3]},
\end{aligned}
\end{equation}

\noindent
if we assume that the emission regions (jet-knots) are $moving$ sources (see a discussion in Stawarz et al. 2004). From the equations (7)$-$(9), we predict the  SSC flux density measured at $\nu_{\rm X}$ to be roughly

\begin{equation}
\begin{aligned}
f_{\rm X} = 2.8 \times 10^{-3} \eta^{-1/2} (1+z) \left( \frac{d_{\rm L}}{\rm 100 Mpc} \right)^{1/2}\\
\times\left( \frac{\nu_{\rm R}}{\rm 5GHz} \right)^{5/4} \left( \frac{\nu_{\rm X}}{\nu_{\rm 1keV}} \right)^{-3/4}\left( \frac{f_{\rm R}}{100{\rm mJy}} \right)^{3/2}\\
\times \left( \frac{\theta}{0.3''} \right)^{-1/2} \delta^{-5/2} \quad {\rm [nJy]}.
\end{aligned}
\end{equation}

\noindent
Here we have assumed $\alpha_{\rm R}$ = 0.75 taking the result of Figure 1 into account. Note that $f_{\rm X}$ goes as $\propto$ $\delta^{-5/2}$, meaning that the SSC flux significantly $decreases$ as the beaming factor increases. Note also that $f_{\rm X}$ $\propto$ $\theta^{-1/2}$, i.e. that for a given $f_{\rm R}$ and $B$ = $B_{\rm eq}$ clumping of the emission region leads to the $increase$ of the SSC X-ray flux.

\vspace{5mm}

\subsection{External Compton (EC) emission on CMB photon field}

Similarly to the SSC case, we can estimate the expected EC flux at a certain X-ray frequency $\nu_{\rm X}$. In the EC model, electrons upscatter CMB photons to frequencies  peaked at $\nu_{\rm IC}$, which, in a Thomson regime, is simply

\begin{equation}
\begin{aligned}
\nu_{\rm IC} \simeq \frac{4}{3}  \gamma_{\rm R}^2 \nu_{\rm CMB} (1+z)^{-1} \delta^2 \kappa = 7.3 \times 10^{18} \eta^{-2/7}\\ 
\times\kappa (1+z)^{-4/7} \left( \frac{d_{\rm L}}{\rm 100 Mpc} \right)^{2/7}\left(\frac{\nu_{\rm R}}{\rm 5GHz} \right)^{6/7}\\
\times\left( \frac{f_{\rm R}}{100 {\rm mJy}} \right)^{-2/7} \left( \frac{\theta}{0.3''} \right)^{6/7}\delta^{12/7} \quad {\rm [Hz]},
\end{aligned}
\end{equation}

\noindent
where $\kappa$ = (1 + $\mu$)(1 + $\beta)^{-1}$ and $\nu_{\rm CMB}$ =
1.6$\times$10$^{11}$(1+$z$) [Hz]. Here we may set $\kappa$ $\simeq$ 1
for simplicity, since the value of $\kappa$ is always an order of unity
for any choice of $\Gamma_{\rm BLK}$  and $\delta$. The ratio of X-ray
(EC) luminosity to the radio (synchrotron) luminosity is approximately
given by (Stawarz et al. 2003)

\begin{equation}
\begin{aligned}
\frac{\nu_{\rm IC} f_{\rm IC}}{\nu_{\rm R} f_{\rm R}}   \simeq \frac{\nu_{\rm X} f_{\rm X}}{\nu_{\rm R} f_{\rm R}} \left( \frac{\nu_{\rm IC}}{\nu_{\rm X}} \right)^{1 - \alpha_{\rm R}} \simeq  \frac{u_{\rm CMB}}{u'_B} \kappa^2 (1+z)^4 \delta^2,
\end{aligned}
\end{equation}

\noindent
where $u_{\rm CMB}$ = 4.1$\times$10$^{-13}$ erg/cm$^3$. From the equations (11) and (12), the EC flux density measured at $\nu_{\rm X}$ can be expressed as

\begin{multline}
\begin{aligned}
f_{\rm X} = 5.9 \times 10^{-4} \kappa^{7/4} \eta^{-1/2} (1+z)
\left( \frac{d_{\rm L}}{\rm 100 Mpc} \right)^{1/2}\\
\times\left( \frac{\nu_{\rm R}}{\rm 5GHz} \right)^{1/2} \left( \frac{\nu_{\rm
 X}}{\nu_{\rm 1keV}} \right)^{-3/4} 
\times\left( \frac{f_{\rm R}}{100{\rm mJy}} \right)^{1/2}\\
\times\left( \frac{\theta}{0.3''} \right)^{3/2} \delta^{3} \quad {\rm [nJy]} .
\end{aligned}
\end{multline}

\noindent
for $\alpha_{\rm R}$ = 0.75. It is interesting to note that $f_{\rm X}$ goes as $\propto$ $\delta^{3}$, meaning that the EC flux significantly $increases$ as the beaming factor increases, which is the exact opposite trend in the SSC case. Note also, that in the case of the EC emission $f_{\rm X}$ $\propto$ $\theta^{3/2}$, i.e. for smaller emission region with given $f_{\rm R}$ and $B$ = $B_{\rm eq}$ the EC X-ray emission decreases, again opposite to what is expected in the case of the SSC process.

\subsection{Source classification}

First we group the sources by the X-ray spectral index $\alpha_{\rm X}$ and the X-ray flux $f_{\rm X}$ observed at 1 keV. If the X-ray emission smoothly connects with the radio/optical spectra, we consider the X-rays to be produced via the synchrotron emission as for the radio to optical photons, and that only the highest energy tail of the electron population contributes to the X-ray emission. Good examples are the knots in M~87, where the X-ray spectral indices are $\alpha_{\rm X}$ $\simeq$ 1.3$-$1.6 and the X-ray fluxes are consistent with radio-optical-X-ray synchrotron continua of a broken power-law form. As listed in Table 3, we find 25 ``synchrotron'' jet-knots and 7 hotspots, but none was found for the radio lobes. Figure 4 plots the distribution of $\gamma_{\rm X}$$\delta^{1/7}$, calculated from the equation (5) derived in $\S$ 3.2. Note that for all the synchrotron sources, electrons must be accelerated very efficiently up to $\gamma_{\rm X}$ $\simeq$ 10$^{7-8}$ for $B$ = $B_{\rm eq}$ (and to even higher energies if only $B$ $<$ $B_{\rm eq}$), and that the highest population is  occupied by the hotspots.

\begin{table}
\begin{center}
Table 3. Source classification of Jets, hotspots, and lobes.

\vspace{3mm}

\begin{tabular}{lccc}
\tableline\tableline
      & Jet-knot & Hotspot & Lobe \\
\tableline
QSO(CD) & 19  & 2 &  0 \\
QSO(LD)  & 7   & 9  & 6  \\
RG(FR I) & 22    &0  & 3 \\
RG(FR II) & 1    &13  & 9 \\
BLZR & 7    &0  & 0 \\
\tableline
SYN & 25  & 7 &  0 \\
SSC  & 4   & 16  & 1  \\
EC & 27    &1  & 17 \\
\tableline
\end{tabular}
\end{center}
\end{table}

\begin{figure}[htb]
\includegraphics[angle=90,scale=.40]{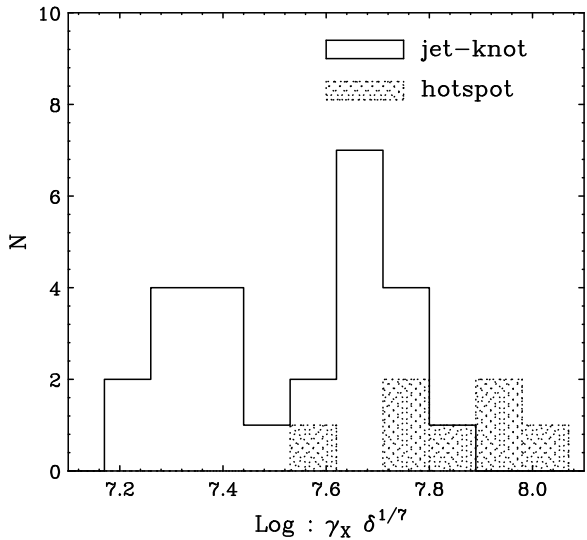}
\caption{Distribution of the electron Lorentz factor, $\gamma_{\rm X}$, for the SYN sources.}
\end{figure}

Meanwhile, remaining sources show flat X-ray spectra which cannot connect smoothly with the radio and optical spectra in terms of single (or broken) power-law continuum. Let us follow the ``conservative'' hypothesis that the X-rays in these sources are due to the inverse-Compton emission of either synchrotron itself (SSC), or the CMB photons (EC). We therefore compare the ratio between the  observed flux density to that expected one from SSC and EC models (c.f., $\S$ 3.3; 3.4),  $R_{\rm SSC}$ and $R_{\rm EC}$,  to determine which process may dominate for the X-ray production. For example, the hotspot of 3C~123 is well explained by SSC, because $R_{\rm SSC}(1)$ = 1.5 and $R_{\rm EC}(1)$ = 140. This means that observed X-ray luminosity is 1.5 times larger than that expected from the SSC model under equipartition hypothesis, whereas 140 times of the expected EC flux if $\delta$ = 1. In contrast, a good example of the EC source are the lobes in 3C~15, where $R_{\rm SSC}(1)$ = 59 and $R_{\rm EC}(1)$ = 1.2. The results of classification are given in the last column of Table 2.

Resultant group of jet-knots, hotspots and radio lobes are summarized in Table 3. Note that most of the jet-knots are either the synchrotron or the EC sources, whereas majority of the hotspots are the SSC sources. Moreover, almost all of the radio lobes emit X-rays via the EC (CMB) process. However, in a number of jet-knots classified as SSC and EC, the observed X-ray luminosities cannot be reproduced satisfactorily. For example, modelling of the jet-knot in PKS~0637 results in $R_{\rm SSC}(1)$ = 600 and $R_{\rm EC}(1)$ = 1,600, meaning that the observed X-ray flux is about 1,000 times brighter than those expected from both the EC and SSC models. As we have derived in $\S$ 3.3 and 3.4, and is well known from the literature, such discrepancy could be reduced by taking the relativistic beaming effect into account, $\delta$ $\neq$ 1, by giving up the equipartition hypothesis, $B$ $<$ $B_{\rm eq}$, or by postulating a synchrotron origin of the X-ray photons due to an additional flat-spectrum electron population. None of these possibilities can be simply excluded. We will comment more about it in the next section.

Let us mention briefly, that due to differences in the model fitting procedure
adopted in this paper as compared to the previous studies reported in the literature,
some differences may occur in either specific values for the obtained model parameters
(e.g., cf. Sambruna et al. 2004 for the case of NGC~6251), or even in classification of
some particular sources (e.g., cf. Fabian et al. 2003 for 3C~9). 
Yet another case is the knot A1 in quasar 3C 273. Marshall et al. (2001) claimed that 
its X-ray emission is consistent with the extrapolation of the radio-to-optical single 
power-law synchrotron continuum. However, Jester et al. (2002) have shown that this is 
not the case, as the observed X-ray flux thereby $exceedes$ the one expected from such 
an extrapolation. Here we classify the 3C~273 knot A1 as the SYN source,
in accordance with
Marshall et al. (2001), although it should be emphazise that --
in face of the detailed optical studies by Jester et al. (2002, 2004) --
this particular choice already involves non-standard energy distribution of the radiating
electrons.

\section{Discussion}

In the previous sections, we have followed ``conservative'' classification of the discussed
sources based on their radio and X-ray emission properties. SYN sources are mainly found as
jet-knots in nearby low-luminosity radio galaxies in agreement with previous studies (e.g.,
Hardcastle et al. 2001a, Pesce et al. 2001, Birkinshaw et al. 2002). 
If the magnetic field strength is not far from the equipartition 
value in these objects, the electrons must be accelerated very efficiently up to 10$-$100 TeV, 
in accordance with the general expectation that radio galaxies may be one of the most efficient 
particle accelerators in the Universe (see a discussion in Kataoka et al. 2003). If the 
electrons are actually accelerated to such high energies, the electrons emitting via synchrotron 
in the X-ray band have relatively short radiative life times. The synchrotron cooling time of 
the electrons can be expressed as

\begin{equation}
t_{\rm sync} = 250 \left( \frac{B_{\rm eq}}{\rm 100 \mu G} \right)^{-2} \left( \frac{\gamma}{\rm 10^7} \right)^{-1} \quad {\rm [yr]}.
\end{equation}
\noindent
Since Comptonization of the synchrotron photons, CMB photons and of the galactic 
photon fields also cool electrons (what can be especially significant if the 
considered jets are relativistic on kpc scales), the above estimate would be an 
upper limit for the electron cooling time scale. Indeed, Stawarz et al. (2003)
estimated the energy density of the starlight emission at 1 kpc from the center of
average elliptical galaxy -- where typically the X-ray bright knots of the low-powerful jets 
are located -- to be $u_{\rm star}$ $\sim$ $10^{-9}$ erg cm$^{-3}$. Now, for the
25 FR I jet-knots classified as SYN sources in this paper the median equipartition magnetic field
computed for non-relativistic bulk velocities is $B_{\rm eq, \, \delta=1}$ = 130 $\mu$G (see Table 2),
what gives the comoving energy density of the magnetic field $u'_{B}$ = 6.7 $\times$ 10$^{-10}$
$\delta^{-10/7}$ erg cm$^{-3}$. Thus, the relative importance of the inverse-Compton to synchrotron
radiative losses for the electrons within the FR I jets is roughly

\begin{equation}
{u'_{\rm star} \over u'_{B}} \sim \Gamma_{\rm BLK}^2 \, \delta^{10/7} \, .
\end{equation}
\noindent
That is, radiating electrons within nearby FR I jets possessing X-ray (and optical) counterparts 
(which are believed to be at least moderately beamed toward the observer, $\delta > 1$), cool mainly 
due to inverse-Compton losses on the starlight photon fields of the host galaxies unless $B$ $\gg$ $B_{\rm eq}$.
Hence, for the highest energy electrons in the FR I jets one can safely put the radiative cooling
spatial scale $ct_{\rm cool}$ $<$ 100 pc. In general, this is consistent with the visual 
sizes of the jet-knots, but significantly smaller than the typical knots' distances from the nucleus 
($\gtrsim$ kpc in the case of FR I sources), and also than the typical inter-knot separation distances. 
Therefore, the jet electrons have to be accelerated $in$ $situ$, most probably due to stochastic processes
connected with strong turbulence occurring within those jets as a result of the mass entrainment from the 
surrounding medium (De Young 1986).

One can ask if in the case of the SYN jet-knots in the nearby FR I galaxies magnetic field can be much smaller than the equipartition value. This possibility could be verified by means of detecting the inverse-Compton radiation of the synchrotron-emitting electrons, which is expected to peak at high-energy $\gamma$-ray band. Interestingly, we can already put some meaningful limits on such high-energy component in the case of the M~87 jet. Nearby radio galaxy M~87 was detected at TeV photon energies by $HEGRA$ Cherenkov Telescope (Aharonian et al. 2003), and it was shown that the kpc-scale jet in this object can produce very high energy $\gamma$-ray photons at the required level via comptonization of the starlight photon field (Stawarz et al. 2003). However, the latest non-detection of M~87 by $Whipple$ Telescope (Le Bohec et al. 2004) suggests variability of the discussed TeV radiation, indicating that the kpc-scale jet in M~87 cannot account for the $HEGRA$ signal. The implied upper limits indicate in turn that the magnetic field within the kpc-scale jet of M~87 radio galaxy cannot be smaller than the equipartition value (Stawarz et al. 2004, in preparation). Thus, one can expect that also in the case of the other FR I jets $B$ $\gtrsim$ $B_{\rm eq}$.

For the SSC and EC sources, a number of jet-knots seem extremely bright in the X-rays, as we have seen in Figure 2 and 3. This inevitably causes a large discrepancy between the ``expected'' and ``observed'' X-ray fluxes as we see in Table 2 and $\S$ 3.5, and what is again well known from the previous studies. One formal possibility is that equipartition hypothesis may not be valid in the considered jet-knots. For a given synchrotron luminosity $L_{\rm sync}$ $\propto$ $u_{\rm e}$$u_B$ and for a given emitting region volume $V$, an expected SSC luminosity is $L_{\rm SSC}$ $\propto$ $u_{\rm e}$. We therefore expect ratio $R_{\rm SSC}$(1) $\propto$ $L_{\rm SSC}^{-1}$ $\propto$ $u_B$. Similarly, for the EC case, $R_{\rm EC}$(1) $\propto$ $L_{\rm EC}^{-1}$ $\propto$ $u_B$. Hence, in both models, the expected X-ray luminosity will be increased by decreasing the magnetic field strength.

\begin{figure}[htb]
\includegraphics[angle=90,scale=.40]{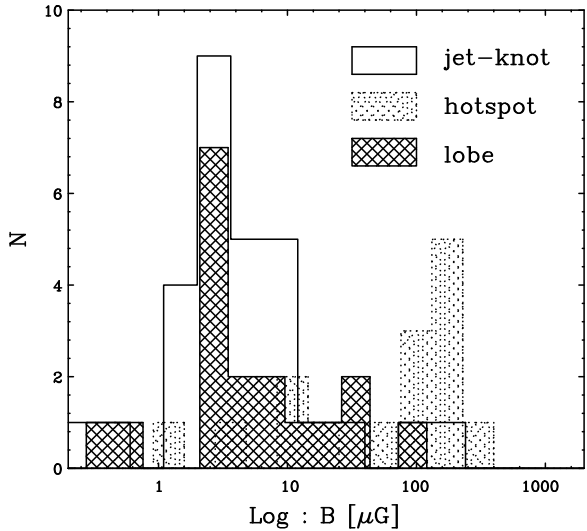}
\caption{Distribution of the evaluated magnetic field, $B$, for the case of $no$ relativistic beaming ($\delta$=1).}
\end{figure}

Figure 5 shows the distribution of the ``best-fit'' magnetic field $B$
if we allow for the deviation from the equipartition condition and
assume nonrelativistic velocities for the emitting regions (what, in the
case of the jet-knots, is rather only a formal hypothesis). One finds
that both the non-SYN jet-knots and radio lobes are distributed around
$B$$\simeq$ 1$-$10$\mu$G, whereas hotspots have a relatively narrow peak
at higher field strength, $B$ $\simeq$ 50$-$300$\mu$G, plus a ``tail''
extending down to $\sim$$\mu$G. Figure 6 shows the ratio of $B$ to the
equipartition value. Interestingly, $B$ in the lobe and most of the
hotspots are almost consistent with the equipartition ($B$/$B_{\rm
eq,\delta=1}$ $\sim$ 1), whereas that of the non-SYN jet-knots and of
some of the hotspots is much weaker from what is expected ($B$/$B_{\rm
eq, \delta=1}$ $\sim$ 0.01$-$0.1).

\begin{figure}[htb]
\includegraphics[angle=90,scale=.40]{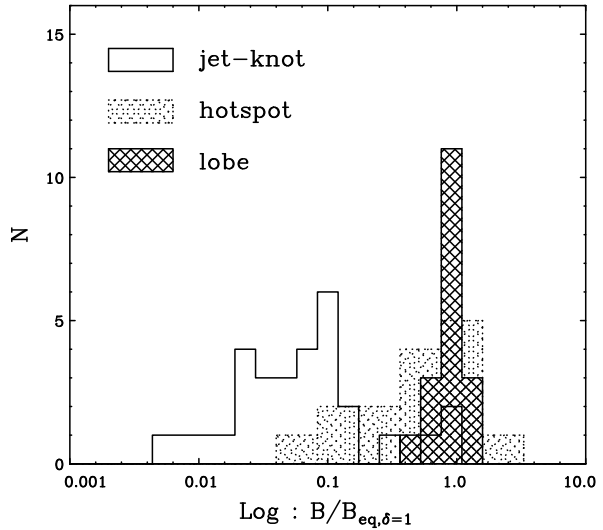}
\caption{Distribution of the ratio between the evaluated magnetic field  $B$ (for $\delta$ = 1) and the equipartition value $B_{\rm eq, \, \delta = 1}$.}
\end{figure}

As an alternative idea, we also consider a case when the difference between the ``expected'' and ``observed'' X-ray fluxes is due to the relativistic beaming effect, and the minimum-power condition is fulfilled, as suggested by Tavecchio et al. (2000) and Celotti et al. (2001). Relativistic beaming changes the observed X-ray luminosities significantly as $f_{\rm X}$ $\propto$ $\delta^{-5/2}$ for the SSC and $\propto$ $\delta^{3}$ for the EC (equation (10) and (13)). Deviation from equipartition may not be formally required so long as an appropriate beaming factor is assumed. The Doppler factors thus calculated are shown in Table 2 and Figure 7. One can see that the lobes and the hotspots exhibit relatively narrow distribution at $\delta$ $\sim$ 1, whereas for most of the jet-knots large beaming factors of $\sim$ 10 are required, as noted before by many authors. We note, that the obtained $\delta$ $\sim$ 0.1 for some of the hotspots is rather a formal possibility. Figure 8 shows the distribution of equipartition magnetic field in the framework of relativistically moving jet model. Similarly to Figure 5, we find again that the narrowly distributed strength of the magnetic field in the hotspots, $B$ $\sim$ 100$-$500$ \mu$G, is an order of magnitude larger than that of the jet-knots and radio lobes.

\begin{figure}[htb]
\includegraphics[angle=90,scale=.40]{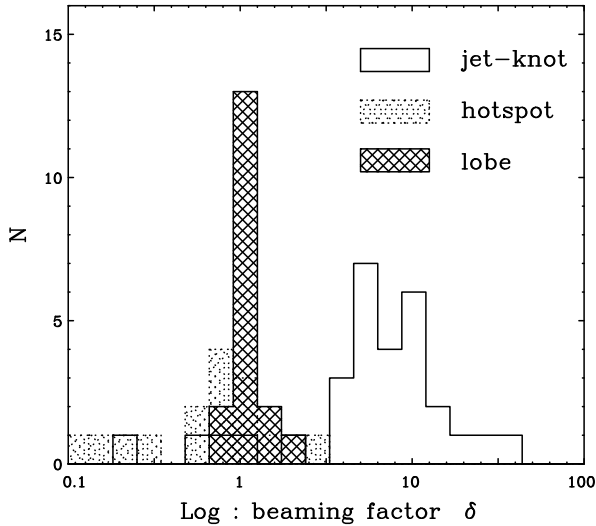}
\caption{Distribution of the required beaming factor $\delta$ for $B$ = $B_{\rm eq}$.}
\end{figure}

\begin{figure}[htb]
\includegraphics[angle=90,scale=.40]{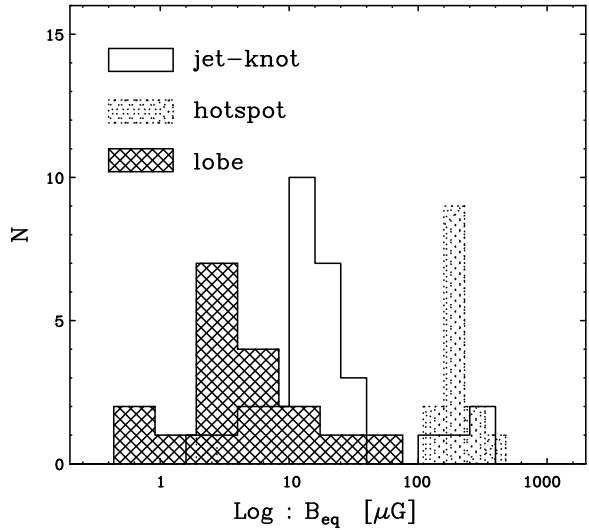}
\caption{Equipartition magnetic field for relativistically moving jet model.}
\end{figure}

Apparently, the above considerations imply that the inverse-Compton X-ray emissions from the lobes and hotspots are relatively close to that expected from the magnetic field--radiating electrons energy equipartition, with at most mildly-relativistic bulk velocities of the radiating plasma. A number of jet-knots in powerful sources requires however significant bulk Lorentz factor of $\Gamma_{\rm BLK}$ $\ge$ $\delta$/2 $\sim$ 5 to agree the inverse-Compton origin of the X-ray photons with the minimum-power condition $B$ = $B_{\rm eq}$. We note, that our evaluation gives the $minimum$ value of $B_{\rm eq}$, as we set $\nu_{\rm min}$ = $\nu$ in the equation (1) and $\eta = 1$ in the subsequent analysis. Therefore, any more realistic derivation would result in an even larger deviation from the energy equipartition, and thus in an even larger value for the jet Doppler factor $\delta$ required to satisfy the minimum power condition. Let us mention, that the alternative two-population synchrotron models do not require violation of the energy equipartition (Stawarz \& Ostrowski 2002, Dermer \& Atoyan 2004).

Usually, in applying the EC model to the quasar jet-knots' X-ray emission, the idea of sub-equipartition magnetic field is rejected since it implies a very high kinetic power of the jets. For this reason, large values for the jet Doppler factors are invoked. However, as discussed by, e.g., Atoyan \& Dermer (2004), such an approach does not solve all the problems with the total energy requirements (see also a discussion in Stawarz 2004). Let us mention in this context another important issue. It is well known, that the $VLA$ studies of the large-scale jets in quasars and FR IIs indicate that bulk Lorentz factors of the radio-emitting plasma in these sources cannot be much greater than $\Gamma_{\rm BLK}$ $\sim$ 3 (Wardle \& Aaron 1997). The discrepancy between this result and the requirement of the minimum-power EC model for $\Gamma_{\rm BLK}$ $>$ 10 is typically ascribed to the jet radial velocity structure, namely that the radio emission originates within slower-moving jet boundary layer and the inverse-Compton X-ray radiation is produced within the fast jet spine (e.g., Ghisellini \& Celotti 2001). While it is true that jet radial stratification can indeed significantly influence jet-counterjet brightness asymmetry ratio, one should be aware that by postulating different sites for the origin of radio and X-ray photons, homogeneous one-zone models for the broad-band knots' emission can no longer be preserved. In particular, in such a case one has to specify exactly what fraction of the jet radio emission is produced within the spine and what fraction within the boundary layer, what is exactly the jet velocity radial profile, and what is the magnetic field strength in each jet components, etc. Without such a discussion one cannot simply use the observed radio flux of the jet to construct the broad-band spectral energy distribution of the knot region, i.e. simply estimate the expected inverse-Compton flux by means of equipartition magnetic field derived from the radio observations. If one insists on applying the homogeneous one-zone model (as a zero-order approximation), as presented in this paper, self-consistency requires a consideration of $\Gamma_{\rm BLK}$ $\leq$ 5. In such a case, a departure from the minimum power conditions within the non-SYN X-ray jets is $inevitable$.

Accordingly to the discussion above, if the X-ray emission of the powerful jets is due to the EC process, these jets are most likely $particle$ $dominated$ ($u_e$$\gg$$u_B$). The jet magnetic field must be then significantly amplified in the hotspot, where an approximate equipartition is expected to be reached. Then the shocked plasma moves slowly to the radio lobe, where the equipartition field becomes close to the intergalactic value ($B$$\sim$ a few $\mu$G). Let us comment in this context on the following issue. Pressure of radio-emitting electrons within the lobes of quasars and FR IIs computed from the equipartition condition is often found to be below the thermal pressure of the ambient medium (Hardcastle \& Worrall 2000), what challenges the standard model for the evolution of powerful radio sources. Such a discrepancy can be removed only by postulating deviations from the equipartition condition, or presence of non-radiating relativistic protons within the lobes. The presented analysis of the X-ray data confirms for a large number of sources the anticipated result that the magnetic field--radiating electrons energy equipartition within the lobes is generally fulfilled, and thus that the relativistic protons are very likely to constitute a significant fraction of the lobes' non-thermal pressure. Interestingly, this would mean that the total energy outputs of powerful jets are systematically $larger$ from what is implied by the analysis of the lobes' radio emission solely (Rawlings \& Saunders 1991). This, in turn, would be consistent with deviation from the minimum-power condition within the considered jets themselves. We note that viscous acceleration of cosmic rays taking place at the turbulent boundary layers of relativistic jets, discussed by, e.g., Ostrowski (2000) and Rieger \& Mannheim (2002), could provide energetically important flat-spectrum population of ultrarelativistic protons within the lobes of powerful radio sources.

We have discussed two different versions of the EC model to account for extremely bright X-ray jet-knots: (1) non-equipartition case and (2) significant relativistic beaming case. Both of these options are in many ways problematic. Our next concern is to attempt to prove in general the postulated inverse-Compton origin of the X-ray photons. Note in this context, that for the EC sources the radio-to-X-ray flux ratio is proportional to $u'_B$$^{-1}$$(1+z)^4$$\delta^2$ (equation 12). Therefore, for a large sample of EC sources one should expect to observe $L_{\rm R}$/$L_{\rm X}$ $\propto$ $(1+z)^4$ behavior, if only $B$ and $\Gamma_{\rm BLK}$ do not have large scatter from source-to-source (cf. Figures 5 and 7). We therefore expect the high-redshift EC sources to be brighter in X-rays than nearby EC sources (see Schwartz 2002, Cheung 2004). 

\begin{figure}[htb]
\includegraphics[angle=90,scale=.33]{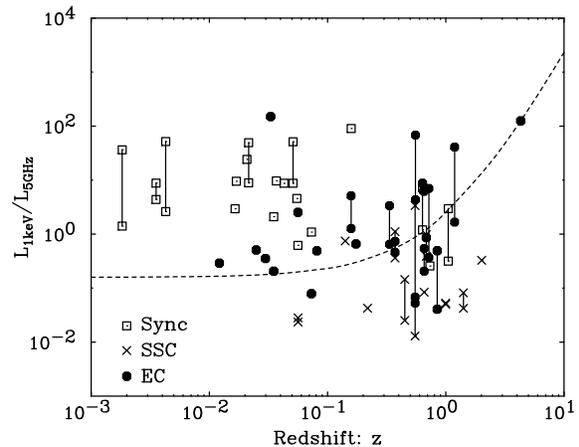}
\caption{Luminosity ratio, $L_{\rm 1keV}$/$L_{\rm 5GHz}$, as a function of redshift for SYN, SSC and EC soures.}
\end{figure}

Figure 9 shows the distribution of the flux ratio ($L_{\rm 1keV}$/$L_{\rm 5GHz}$) as a function of $z$. The dotted line shows $\propto$ $(1+z)^4$ relation which fits the highest $z$ data point (GB~1508+5714; $z$ = 4.3) just to help guide the eyes. Although data sample is still poor, we may say that no clear trend can be seen in this plot. Furthermore, we notice that the discussed ratio is widely distributed even in the same objects. For example, in cases of knots in 4C~19.33 ($z$ = 0.72), ``conservatively'' classified as the EC sources, the X-ray-to-radio luminosity ratio changes of about an order of magnitude (Table 2). Such a difference is not easy to explain in the framework of model (1), since we have to assume an order of magnitude increase in the magnetic fields along the jet. In a framework of relativistic beaming hypothesis (2) one may possibly explain such variation by postulating the decrease of the bulk Lorentz factor along the flow and only moderate changes in magnetic field (Georganopoulos \& Kazanas 2004). In this case, however, one has to explain what causes significant deceleration of the jet, which preserves its excellent collimation, with no significant radiative energy losses. We need more data obtained in various energy bands, and a more sophisticated analysis to conclude this further. However, we note that recent observations of high-redshift quasars by Bassett et al. (2004) did not reveal any evidence for the enhanced X-ray emission of the distant large-scale jets due to the increased energy density of the cosmic microwave background. Since such an effect is expected in a framework of the EC model, one may conclude that the X-ray photons from the powerful quasar jets are not inverse-Compton in origin. Recent detailed re-analysis of the $Chandra$ data for 3C~120, again ``conservatively'' classified as an EC source, strongly support this idea (Harris et al. 2004).

\begin{figure}[htb]
\includegraphics[angle=90,scale=.33]{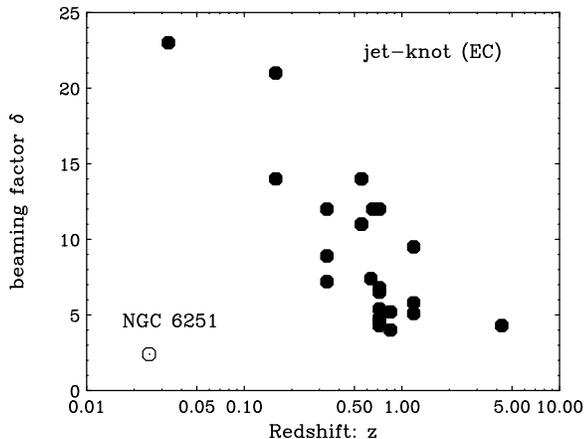}
\caption{Expected beaming factor $\delta_{\rm EC}$ for $B$ = $B_{\rm eq}$, 
as a function of redshift for EC jet-knot sources. $Open$ $circle$ shows
 FR~I radio galaxy NGC~6251, but its X-ray emission seems to be
 problematic in a framework of the EC model. Full details are given in
 the text.}
\end{figure}

Let us finally discuss yet another issue regarding the EC scenario for 
the quasar jets' X-ray emission. Figure 10 shows the Doppler beaming factor $\delta$ required
in this model to obtain $B$ $=$ $B_{\rm eq}$, versus the redshifts $z$ of the jet-knots classified 
here as the EC ones. One can clearly see a significant anticorrelation between $\delta$ and $z$,
meaning that the high-$z$ sources require much smaller $\delta$ for the magnetic field-radiating electrons
energy equipartition than the sources located closer to the
observer.\footnote{Nearby FR I radio galaxy NGC~6251 ($open$ $circle$) constitutes the only exception from this trend. However, this peculiar source does not belong to 
the quasar class, and, in general, its X-ray emission is particularly problematic (especially in 
a framework of the EC model).} There are two possible explanations for the noted $\delta$--$z$
anticorrelation. If reflecting physical property, it would mean that the distant large-scale quasar jets
are less relativistic than their nearby analogues but similarly close to the equipartition, $or$ that
both low- and high-$z$ quasar jets are only mildly relativistic on large scales but closer to the 
minimum-power condition when located at large redshifts. None of this options appear to be particularly
inartificial, especially as the high-$z$ quasar cores seem to be comparable to their low-$z$ counterparts (e.g., 
Bassett et al. 2004). On the other hand, differences in velocity and energy content of the large-scale jets may
not reflect differences in the central engines, but more likely differences in the surrounding galactic or
intergalactic medium. The second possibility for understanding $\delta$--$z$ anticorrelation is however that
it is simply an artifact of the applied but inappropriate EC model. This issue has to be discussed carefully
for a larger number of sources.

\section{Conclusion}

We have studied the statistical properties of the large-scale jet-knots,
hotspots and lobes in more than 40 radio galaxies recently observed with
$Chandra$ and $ASCA$. For the jet-knots in nearby low-luminosity radio
galaxies and for some of the hotspots, X-ray photons are most likely
synchrotron in origin, being then produced by ultrarelativistic
electrons with energies 10$-$100 TeV that must be accelerated in
situ. For the other objects X-ray photons are inverse-Compton in origin,
or, alternatively, are due to synchrotron emission of very high energy
electrons with a non-standard energy distribution. In this paper we
examine in more detail the former possibility. We
first calculated the ``expected'' SSC or EC fluxes by assuming
equipartition magnetic field and nonrelativistic velocity of the
emitting plasma, and then compared them to the observed fluxes. 
We confirmed that the observed X-ray fluxes from the hotspots and 
radio lobes are approximately consistent with the expected ones, 
whereas a number of the jet-knots in powerful sources is too bright 
at X-rays. We examined two possibilities to account for this discrepancy 
in a framework of the inverse-Compton model. The first idea is that 
equipartition hypothesis may not be valid for the considered sources. 
In this case, the X-ray bright jets are particle dominated and therefore 
far from the minimum-power condition. The jet magnetic field must be 
then significantly amplified in the hotspots where an approximate energy 
equipartition with the radiating particles is expected to be reached. 
An alternative idea is that the jets are highly relativistic 
($\Gamma_{\rm BLK}$ $\ge$ 5) even on kpc/Mpc scales, but significantly 
decelerate in the hotspots. This however, in addition to other problems, 
challenges the homogeneous one-zone emission region model adopted in this 
paper, as discussed in the text. Unfortunately, the comparison of the 
observed radio-to-X-ray flux ratios for various $z$ sources 
\emph{from the compiled dataset} does not provide definite constraints 
on the X-ray emission process dominating within the quasar and FR II jets.

\acknowledgments

We would like to thank F. Takahara and M. Ostrowski for fruitful
discussion and constructive comments. J.K. acknowledges a support by 
JSPS.KAKENHI (14340061). \L .S. was supported by the grant PBZ-KBN-054/P03/2001
and partly by the Chandra grants G02-3148A and G0-09280.01-A.

\begin{deluxetable}{lllll}
\tabletypesize{\scriptsize}
\tablecaption{List of Radio Sources with Extended X-ray Jet, Hotspot, and Lobe Structures}
\tablewidth{0pt}
\tablehead{
\colhead{name} & \colhead{$z$} 
& \colhead{$d_{\rm L}$[Mpc]$^a$}
& \colhead{class$^b$} & \colhead{reference}
}
\startdata
3C~9 & 2.012 & 16133 & QSO(LD) & Bridle et al. 1994, Fabian, Celotti \& Johnstone 2003 \\
3C~15 & 0.073 & 302 & RG(FR I) & Leahy et al. 1997; Kataoka et al. 2003b \\
NGC315 & 0.0165 & 67 & RG(FR I) & Worrall, Birkinshaw \& Hardcastle  2003\\
3C~31 & 0.0169 & 67 & RG(FR I) & Laing \& Bridle 2002; Hardcastle et
 al. 2002a\\
NGC~612 & 0.0298 & 120 & RG(FR II) & Isobe 2002 \\
B0206+35 & 0.0369 & 150 & RG(FR I) & Worral, Birkinshaw \& Hardcastle  2001  \\
3C66B & 0.0215 & 87 & RG(FR I) & Hardcastle, Birkinshaw \& Worrall 2001a \\
Fornax~A & 0.00587 & 23.5 & RG(FR I) & Tashiro et al. 2001\\
3C~120 & 0.033 & 134 & RG(FR II) & Harris et al. 1999; 2004\\
3C~123 & 0.218 & 965 & RG(FR II) &  Hardcastle et al. 1997; Hardcastle, Birkinshaw, \& Worrall, 2001b\\
3C~129 & 0.0208 & 84 & RG(FR I)  &  Harris, Krawczynski \& Taylor 2002 \\
Pictor~A & 0.035 & 143 & RG(FR II)  & Wilson, Young \& Shopbell 2001; Isobe 2002 \\
PKS0521-365 & 0.055 & 225 & BLZR  &  Birkinshaw, Warrall \&
 Hardcastle 2002 \\
PKS0637-752 & 0.653 & 3465 & BLZR & Chartas et al. 2000; Schwartz et al. 2000\\
3C~179 & 0.846 & 4815 & QSO(LD)  &   Sambruna et al. 2002\\
B2~0738+313 & 0.635 & 3344 & QSO(CD)  &  Siemiginowska et al. 2003a \\
B2~0755+37 & 0.0428 & 175 & RG(FR I)  &  Worrall Birkinshaw \& Hardcastle 2001 \\
3C~207 & 0.684 & 3642 & QSO(LD)  & Brunetti et al. 2002 \\
3C~212 & 1.049 & 6393 & QSO(LD)  & Akujor et al. 1991, Aldcroft et al. 2003 \\
3C~219 & 0.174 & 756 & RG(FR II)  &  Comastri et al. 2003 \\
4C73.08 & 0.0581 & 236 & RG(FR II)  &   Isobe 2002 \\
Q0957+561 & 1.41 & 9613 & QSO(CD)  & Harvanek et al.1997; Chartas et al.2002 \\
3C~254 & 0.734 & 4011 & QSO(LD) &  Donahue, Daly \& Horner 2003 \\
PKS1127-145 & 1.187 & 7505 & QSO(CD) &   Siemiginowska et al. 2002 \\
PKS1136-135 & 0.554 & 2830 & QSO(LD)  & Sambruna et al. 2002 \\
3C~263 & 0.656 & 3487 & QSO(LD)  &  Hardcastle et al. 2002b \\
4C~49.22 & 0.334 & 1559 & QSO(CD)  & Sambruna et al. 2002  \\
M~84 & 0.00354 & 11.3 & RG(FR I)  & Harris et al. 2002 \\
3C~273 & 0.1583 & 683 &  BLZR & Marshall et al. 2001; Sambruna et al.
 2001\\
M87 & 0.00427 & 10.7 & RG(FR I) & Marshall et al. 2002; Wilson \& Yang
 2002; Perlman et al. 2001\\
3C~280 & 0.996 & 5964 & RG(FR II)  &  Donahue, Daly \& Horner 2003 \\
Cen~A & 0.00183 & 2.3 & RG(FR I) &   Kraft et al. 2002 \\
Cen B & 0.01215 & 48.7 & RG(FR I) & Tashiro et al. 1998\\
4C19.44 & 0.720 & 3917 & QSO(CD)  &  Sambruna et al. 2002\\
3C~295 & 0.45 & 2205 & RG(FR II)  &  Harris et al. 2000: Brunetti et al. 2001a \\
3C~303 & 0.141 & 603 & RG(FR II)  &  Meisenheimer, Yates \& R{\oe}ser 1997; 
 Kataoka et al. 2003a \\
GB1508+5714 & 4.3 & 54142 & QSO(CD)  &  Siemiginowska et al.2003b; Yuan et
 al. 2003; Cheung 2004\\
3C~330 & 0.55 & 2803 & RG(FR II)  &   Hardcastle et al. 2002b\\
NGC6251 & 0.0249 & 101 & RG(FR I)  & Sambruna et al.2004  \\
3C~351 & 0.372 & 1763 & QSO(LD)  &  Brunetti et al.2001b; Hardcastle et al. 2002b \\
3C~371  & 0.051 & 209 & BLZR &  Pesce et al. 2001  \\
3C~390.3 & 0.0561 & 230 &  RG(FR II) &  Harris, Leighly \& Leahy 1998; 
this work \\
Cyg A & 0.0562 & 231 &  RG(FR II) & Wilson, Young \& Shopbell 2000 \\
3C~452 & 0.0811 & 331 & RG(FR II)  &  Isobe et al. 2002 \\
\enddata

\tablenotetext{a}{$d_{\rm L}$; Luminosity distance to the source adopting  $H_0$ = 75 km
 s$^{-1}$ Mpc$^{-1}$ and $q_0$ = 0.5.}
\tablenotetext{b}{RG; radio galaxy of either Fanaroff \& Riley type I
 (FR I) or type II (FR II), QSO: quasar of either core dominated (CD) or
 lobe dominated (LD), and BLZR: blazars.}

\end{deluxetable}

\begin{deluxetable}{llllllllllllll}
\tabletypesize{\scriptsize}
\tablecaption{Parameters for X-ray Jet, Hotspot, and Lobe Features}
\tablewidth{0pt}
\tablehead{
\colhead{name} 
& \colhead{comp}
& \colhead{$\alpha_{\rm R}$}
& \colhead{$f_{\rm R}$}
& \colhead{$\alpha_{\rm X}$}
& \colhead{$f_{\rm X}$}
& \colhead{$f_{\rm O}$}
& \colhead{$\theta$}
& \colhead{$B_{{\rm eq}}(1)$}
& \colhead{$R_{{\rm SSC}}(1)$}
& \colhead{$R_{{\rm EC}}(1)$}
& \colhead{$\delta_{\rm SSC}$}
& \colhead{$\delta_{\rm EC}$}
& \colhead{class}\\  &  & & [mJy] &  & [nJy] & [$\mu$Jy]  & [''] & [$\mu$G]
 & &  & & 
}
\startdata
3C9 & K & 1.0f & 3.2e2  &  0.6$\pm$0.6 &  2.2 & --  & 0.3f
 & 230 & 3.6 & 54 & {\bf 0.60} & 3.8  &  SSC \\
3C15 & K-C & 0.9 & 55  &  0.7$\pm$0.4 &  1.2 &  6.0 & 0.3f
 & 84 & 5.5e2 & 1.5e3 & 0.080 & 11  &  SYN \\
      & L & 0.8 & 1.8e3  &  0.3$\pm$0.4 &  2.9 &  --  & 20
 & 6.2 & 58 & 1.1 & 0.20 & {\bf 1.0}  &  EC \\
NGC315 & K & 0.9f & 68  &  1.5$\pm$0.7 &  4.1 &  --   & 0.3f
 & 130 & 3.1e3 & 1.0e4 & 0.040 & 22  &  SYN \\
3C31 & K & 0.55 & 37  &  1.1$\pm$0.2 &  7.3 &   2.0  & 0.3f
 & 110 & 1.4e4 & 2.4e4 & 0.022 & 29  &  SYN \\
NGC612 & L & 0.6 &  5.1e3 & 1.0$\pm$0.5 &  37  &  --   & 120
 & 2.2 & 6.4e2 & 0.99 & 0.075 & {\bf 0.99}  &  EC \\
B0206 & K & 0.5 & 26  &  1.1$\pm$0.7 &  5.2 &    --   &  0.3f
 & 79 & 1.1e4 & 1.4e4 & 0.024 & 24  &  SYN \\
3C66B & K-A & 0.75 & 3.9  &  1.0$\pm$0.3 &  4.0 &  1.0 & 0.3f
 & 52 & 1.9e5 & 3.6e4 & 7.7e-3 & 33  &  SYN \\
      & K-B & 0.60 & 34  &  1.2$\pm$0.1 &  6.2 &  15.8  & 0.3f
 & 97 & 1.2e4 & 1.9e4 & 0.024 & 27  &  SYN \\
For~A & L & 0.9 & 1.6e4 & 0.7$\pm$0.3 & 100 &    --    & 450
 & 1.5 & 1.4e3 & 0.47 & 0.055 & {\bf 0.78}  &  EC \\
3C120 & K & 0.65 & 9.2  &  0.5$\pm$0.1 & 29 &  $<$0.7   & 1.5
 & 15 & 6.9e5 & 1.2e4 & 4.6e-3 & {\bf 23}  &  EC \\
3C123 & HS & 0.5 & 5.2e3  &  0.6$\pm$0.3 & 4.6  & $<$3   & 0.5
 & 170 & 1.5 & 1.3e2 & {\bf 0.85} & 5.1  &  SSC \\
3C129 & K & 0.5f & 3.8  &  1.0f & 1.9 &   --    & 0.3f 
 & 52 & 9.7e4 & 1.8e4 & 0.010 & 26  &  SYN \\
Pic~A& HS & 0.74 & 2.0e3  &  1.1$\pm$0.1 & 87 & 104     & 0.5
 & 180 & 3.6e2 & 1.2e4 & 0.095 & 23  &  SYN \\
& L-W & 0.72 & 1.3e4 & 0.6$\pm$0.3 & 56 &   --  & 90
 & 3.5 & 1.9e2 & 1.3 & 0.12 & {\bf 1.1}  &   EC \\
PKS0521& K & 0.6 & 1.5e2  &  1.3$\pm$0.3 & 14 &  45  &  0.3f
 & 120 & 1.7e3 & 1.2e4 & 0.050 & 23  &  SYN \\
PKS0637& K & 0.8 & 48  &  0.9$\pm$0.1 & 6.2 & 0.2   & 0.3f
 & 80 & 6.0e2 & 1.6e3 & 0.077 & {\bf 12}  &  EC \\
3C179& K-A & 0.8 & 73  &  1.0f & 0.40 & $<$0.06    &  0.3f
 & 97 & 15 & 61 & 0.34 & {\bf 4.0}  &  EC \\
      & K-B & 0.8 & 1.1e2 &  1.0f & 1.1 & $<$0.06 & 0.3f
 & 110 & 23 & 1.4e2 & 0.29 & {\bf 5.2}  &  EC \\
      & CL & 0.8f & 2.9e2 &  1.0f & 0.24 & $<$2.8 & 2.0
 & 28 & 3.0 & 1.1 & 0.64 & {\bf 1.0}  &  EC \\
B2~0738& K-A & 0.5f & 1.7  &  0.5$\pm$0.4 & 0.30 & --   & 0.3f
 & 30 & 4.5e3 & 4.1e2 & 0.035 & {\bf 7.4}  &  EC \\
    & HS-B1 & 0.5f & 2.2  &  1.0$\pm$0.3 & 0.33 & -- &  0.3f
 & 33 & 3.4e3 & 4.0e2 & 0.039 & 7.4  &  SYN \\
    & HS-B2 & 0.5f & 4.0  &  1.4$\pm$0.5 & 0.10 & -- & 0.3f
 & 39 & 4.2e2 & 89 & 0.090 & 4.5  &  SYN \\
B2~0755 & K & 0.5f & 54  &  1.1$\pm$0.2 & 9.7 & -- & 0.3f
 & 94 & 6.3e3 & 1.6e4 & 0.030 & 25  &  SYN \\
3C207 & K & 0.8 & 2.3e2  &  0.2$\pm$0.3 & 4.6 & -- & 0.3f
 & 130 & 46 & 5.1e2 & {\bf 0.22} & 8.0  &  SSC \\
 &HS &  0.8 & 1.6e2  &  0.7$\pm$1.0 & 1.3 & -- & 0.3f
 & 110 & 23 & 1.7e2 & {\bf 0.29} & 5.6  &  SSC \\
 & L & 0.9 & 2.5e2  &  0.5$\pm$0.4 & 4.5 & -- & 5.0
 & 12 & 1.4e2 & 7.0 & 0.14 & {\bf 1.9}  &  EC \\
3C212 &HS-S? & 0.5f & 13  &  1.0f & 0.80 & -- & 0.3f
 & 64 & 3.1e2 & 2.3e2 & 0.10 & 6.1  &  SYN \\
       &HS-N? & 0.5f & 74  &  1.0f & 0.48 & -- & 0.3f
 & 106 & 14 & 58 & 0.35 & 3.9  &  SYN \\
3C219 & L & 0.8 & 2.2e3  &  0.7$\pm$0.2 & 30 & -- & 50
 & 2.7 & 4.0e2 & 1.6 & 0.091 & {\bf 1.2}  &  EC \\
4C73.08 & L-E & 0.85 & 2.7e2 & 0.7$\pm$0.4 & 54 & -- & 180
 & 0.58 & 6.4e4 & 2.3 & 0.012 & {\bf 1.3}  &   EC \\
        & L-W & 0.85 & 5.6e2 & 0.65f & 31 & -- & 180
 & 0.71 & 1.2e4 & 0.93 & 0.023 & {\bf 0.98}  &   EC \\
Q0957 & K-B & 0.8f & 2.2e2  &  0.9$\pm$0.6 & 0.37 & $<$0.11 & 0.3f
 & 170 & 1.7 & 18 & {\bf 0.81} & 2.6  &  SSC \\
         & K-C & 0.8f & 1.3e2  &  0.9$\pm$0.6 & 0.11 & $<$0.11 & 0.3f
 & 140 & 1.1 & 6.9 & {\bf 0.96} & 1.9 &  SSC \\
3C254 & HS-W & 0.8f & 98 &  1.0$\pm$0.8 & 0.52 & -- & 0.3f
 & 100 & 15 & 81 & 0.34 & 4.3  &  SYN \\
PKS1127 & K-A & 1.2 & 1.3 &  0.5f & 1.1 & -- & 0.3f
 & 35 & 1.1e4 & 8.6e2 & 0.024 & {\bf 9.5}  &  EC \\
            & K-B & 0.82 & 16 &  0.5f & 0.89 & $<$0.18 & 0.3f
 & 72 & 2.2e2 & 2.0e2 & 0.11 & {\bf 5.8}  &  EC \\
            & K-C & 0.86 & 17 &  0.5f & 0.60 & $<$0.15 & 0.3f
 & 73 & 1.3e2 & 1.3e2 & 0.14 & {\bf 5.1}  &  EC \\
PKS1136 & K-A & 0.8f & 1.0 &  1.0f & 1.41 & 0.23  & 0.3f
 & 25 & 5.4e4 & 2.9e3 & 0.013 & {\bf 14}  &  EC \\
            & K-B & 0.8f & 41 &  1.0f & 3.7 & 0.24 & 0.3f
 & 73 & 5.4e2 & 1.2e3 & 0.081 & {\bf 11}  &  EC \\
            & K-C & 0.8f & 1.9e2 &  1.0f & $<$0.62 & 0.13 & 0.3f
 & 110 & $<$9.1 & $<$92 & $>$0.41 & {\bf $<$4.5}  &  EC \\
3C263 & HS-K & 0.8f & 5.7e2 &  1.0$\pm$0.1 & 1.0 & 0.8 & 0.3f
 & 160 & 2.7 & 72 & {\bf 0.67} & 4.2  &  SSC \\
       & HS-B & 0.8f & 22 &  1.0f & $<$0.06  & --  &  0.3f
 & 64 & $<$19 & $<$22 & {\bf $>$0.31} & $<$2.8  &  SSC \\
       & L-NW & 0.8f & 1.9e2 &  0.4$\pm$0.2 & 0.8 & --  & 8
 & 7.1 & 51 & 0.73 & 0.21 & {\bf 0.90}  &  EC \\
       & L-SW & 0.8f & 44 &  0.4$\pm$0.2 & 0.5 & --  & 8
 & 4.7 & 2.8e2 & 0.95 & 0.10 & {\bf 0.98}  &  EC \\
4C49.22 & K-A & 0.8f & 56 &  1.0f & 3.9 & 0.63 & 0.3f 
 & 75 & 5.9e2 & 1.7e3 & 0.08 & {\bf 12}  &  EC \\
         & K-B & 0.8f & 36 &  1.0f & 1.3 & 0.02 & 0.3f
 & 66 & 3.8e2 & 7.0e2 & 0.09 & {\bf 8.9}  &  EC \\
         & K-C & 0.8f & 74 &  1.0f & 0.99 & 0.08 & 0.3f
 & 81 & 98 & 3.7e2 & 0.16 & {\bf 7.2}  &  EC \\
M84 & K-2.5 & 0.65 & 3.5 & 0.8$\pm$0.3 & 0.63 & $<$30 & 0.3f
 & 88 & 1.0e5 & 1.7e4 & 9.9e-3 & 26  &  SYN \\
     & K-3.3 & 0.65 & 13 & 0.8$\pm$0.3 & 1.16 & $<$30  & 0.3f
 & 130 & 2.6e4 & 1.6e4 & 0.02 & 25  &  SYN \\
3C273 & K-A1 & 0.65 & 20 & 0.6$\pm$0.1 & 38.1 & 5.2 & 0.3f
 & 56 & 4.8e4 & 4.8e4 & 0.01 & 36  & SYN  \\
       & K-B1 & 0.65 & 2.2e2 & 0.9$\pm$0.1 & 23.2 & 5.2  & 0.3f
 & 110 & 8.1e2 & 8.7e3 & 0.069 & {\bf 21}  &  EC \\
       & K-D/H3 & 0.65 & 3.2e2 & 0.8$\pm$0.1 & 8.27 & 8.2 & 0.3f
 & 120 & 1.6e2 & 2.6e3 & 0.13 & {\bf 14}  &  EC \\
M87 & K-HST1 & 0.7 & 77 & 1.3$\pm$0.1 & 81.9 & 20  & 0.3f
 & 220 & 1.3e5 & 4.8e5 & 9.0e-3 & 78  &  SYN \\
    & K-A & 0.7 & 3.5e2 & 1.3$\pm$0.2 & 67.8 & 100  & 0.3f
 & 330 & 1.1e4 & 1.9e5 & 0.024 & 57  &  SYN \\
    & K-D & 0.7 & 2.6e3 & 1.6$\pm$0.1 & 142 & 1000 & 0.3f
 & 590 & 1.2e3 & 1.4e5 & 0.059 & 52  &  SYN \\
3C280 & HS-W & 0.8 & 7.2e2 & 1.3$\pm$1.0 & 0.79 & 0.99 & 0.3f
 & 200 & 0.95 & 32 & {\bf 1.0} & 3.2  &  SSC \\
       & HS-E & 0.8 & 3.3e2 & 1.2 & 0.34 & 0.23  & 0.3f
 & 160 & 1.3 & 21 & {\bf 0.90} & 2.7  &  SSC \\
Cen~A & K-NX1 & 0.8f & 36 & 1.5 & 5.8 & --  &0.3f
 & 270 & 6.3e4 & 1.1e5 & 0.01 & 48  &  SYN \\
      & K-AX1 & 0.8f & 5.2e2 & 1.5 & 110 & --  &0.3f
 & 580 & 2.2e4 & 5.4e5 & 0.02 & 81  &  SYN \\
      & K-AX2 & 0.8f & 4.8e2 & 1.5 & 14 & --  &0.3f
 & 570 & 3.1e3 & 7.1e4 & 0.04 & 41  &  SYN \\
      & K-AX3 &  0.8f & 7.4e2 & 1.2 & 28 & --  &0.3f
 & 640 & 3.3e3 & 1.1e5 & 0.04 & 49  &  SYN \\
      & K-AX4 &  0.8f & 3.3e2 & 1.2 & 14 & --  &0.3f
 & 510 & 5.5e3 & 8.6e4 & 0.03 & 44  &  SYN \\
      & K-AX6 & 0.8f & 71 & 1.2 & 23 & --  &0.3f
 & 330 & 9.0e4 & 3.0e5 & 0.01 & 67  &  SYN \\
      & K-BX2 & 0.8f & 88 & 1.0 & 66 & --  &0.3f
 & 350 & 1.9e5 & 7.8e5 & 7.8e-3 & 92  &  SYN \\
      & K-BX5 & 0.8f & 7.5e2 & 1.0 & 50 & --  &0.3f
 & 640 & 5.7e3 & 2.0e5 & 0.03 & 59  &  SYN \\
Cen~B & L & 0.78 & 3.7e4 & 0.9$\pm$0.2 & 220 & -- & 180 
 & 3.5 & 3.8e2 & 1.9 & 0.09 & {\bf 1.2}  &   EC \\
4C19.44 & K-A & 0.8f & 57 & 1.0f & 8.3 & 0.3  &0.3f
 & 86 & 5.6e2 & 1.7e3 & 0.08 & {\bf 12}  &  EC \\
        & K-B & 0.8f & 23 & 1.0f & 0.24 & 0.04 &0.3f
 & 66 & 63 & 79 & 0.19 & {\bf 4.3}  & EC \\
        & K-C & 0.8f & 13 & 1.0f & 0.37 & $<$0.06 &0.3f
 & 56 & 2.3e2 & 1.6e2 & 0.11 & {\bf 5.4}  &  EC \\
        & K-D & 0.8f & 16 & 1.0f & 0.25 & $<$0.06  &0.3f
 & 60 & 1.1e2 & 98 & 0.15 & {\bf 4.6}  &  EC \\
        & K-E & 0.8f & 6 & 1.0f & 0.25 & $<$0.06  &0.3f
 & 45 & 4.9e2 & 1.6e2 & 0.08 & {\bf 5.4}  &  EC \\
        & K-F & 0.8f & 12 & 1.0f & 0.70 & $<$0.06  &0.3f
 & 55 & 4.9e2 & 3.2e2 & 0.08 & {\bf 6.8}  &  EC \\
        & K-G & 0.8f & 13 & 1.0f & 0.62 & $<$0.06  &0.3f
 & 56 & 3.8e2 & 2.7e2 & 0.09 & {\bf 6.5}  &  EC \\
        & K-H & 0.8f & $>$1 & 1.0f & 0.41 & $<$0.06 &0.3f
 & $>$27& $>$1.2e4 & $>$6.4e2 & $>$0.02 & {\bf $<$8.6}  &  EC \\
        & K-I & 0.8f & 87 & 1.0f & 0.66 & --  &0.3f
 & 97 & 24 & 1.1e2 & 0.28 & {\bf 4.8}  &  EC \\
3C295 & HS-NW & 0.65 & 1.3e3 & 0.9$\pm$0.5 & 3.9 & 0.078 &0.3f
 & 190 & 4.4 & 2.7e2 & {\bf 0.55} & 6.5  &  SSC \\
       & HS-SE & 0.65 & 6.3e2 & 0.9$\pm$0.5 & 1.1 & $<$0.02 &0.3f
 & 150 & 3.6 & 1.1e2 & {\bf 0.60} & 4.8  &  SSC \\
       & L & 0.9 & 6.5e3 & 0.4$\pm$0.2 & 3.4 & --  &1.5
 & 75 & 0.76 & 9.4 & {\bf 1.1} & 2.1  &  SSC \\
3C303 & HS & 0.84 & 2.6e2 & 0.4$\pm$0.2 & 4.0 & 7.5 & 1.0
 & 42 & 2.2e2 & 2.5e2 & {\bf 0.12} & 6.3  &  SSC \\
GB1508 & K & $>$0.8 & 0.43 & 0.9$\pm$0.4 & 1.1 & $<$0.2 & 0.6
 & 32 & 1.6e4 & 81 & 0.02 & {\bf 4.3}  &  EC \\
3C330 & HS-NE & 1.0 & 1.3e3 & 0.5f & 0.34 & $<$0.5 & 0.3f
 & 200 & 0.32 & 19 & {\bf 1.6} & 2.7  &  SSC \\
       & HS-SW & 1.0 & 1.3e2 & 0.5f & 0.09 & $<$0.5  & 0.3f
 & 100 & 2.6 & 16 & {\bf 0.68} & 2.5  &  SSC \\
      & L-NE & 0.9 & 2.6e2 & 0.5f & 0.28 & -- & 3.5
 & 15 & 8.9 & 0.90 & 0.42 & {\bf 1.0}  &  EC \\
      & L-SW & 1.0 & 2.3e2 & 0.5f & 0.32 & -- & 3.5
 & 15 & 12 & 1.1 & 0.37 & {\bf 1.0}  &  EC \\
NGC6251 & K & 0.64 & 2.2e2 & 0.2$\pm$0.4 & 2.3 & -- & 10
 & 7.9 & 1.4e3 & 13 & 0.06 & {\bf 2.4}  &  EC \\
3C351 & HS-J & 0.7 & 1.9e2 & 0.5$\pm$0.1 & 4.3  & 2.5 & 0.3f
 & 110 & 1.0e2 & 9.2e2 & {\bf 0.16} & 9.7  &   SSC \\
        & HS-L & 0.7 & 4.5e2 & 0.9$\pm$0.1 & 3.4 & 3.8 & 0.8
 & 59 & 36 & 1.1e2 & {\bf 0.24} & 4.8  &  SSC \\
        & HS-A & 0.8 & 4.5 & 0.9f & $<$0.05 & -- & 0.3f
 & 37 & $<$3.0e2 & $<$69 & {\bf $>$0.10} & $<$4.1  &  SSC \\
        & L-N & 1.0 & 72 & 0.6$\pm$0.8 & 1.1 & -- & 10
 & 4.0 & 5.9e2 & 2.0 & 0.08 & {\bf 1.3}  &  EC \\
        & L-S & 0.9 & 73 & 0.6$\pm$0.8 & 0.7 & --  & 10
 & 4.0 & 3.7e2 & 1.3 & 0.09 & {\bf 1.1}  &  EC \\
3C371  & K-A & 0.76 & 37 & 1.0f & 6.7 & 5.8 & 0.3f
 & 81 & 6.9e3 & 1.2e4 & 0.03 & 23  &   SYN \\
        & K-B & 0.73 & 15 & 0.7$\pm$0.3 & 16 & 3.4 & 0.3f
 & 62 & 6.4e4 & 4.6e4 & 0.01 & 36  &   SYN \\
3C390.3 & HS-NE-B & 0.7 & 3.5e2 & 0.9$\pm$0.1 & 4.5 & 1.8 & 1.1
 & 49 & 2.9e2 & 3.6e2 & 0.10 & 7.1  &   SYN \\
        & HS-SW & 0.7 & 67 & 0.4$\pm$0.2 & 3.5 & -- & 10
 & 4.7 & 8.1e3 & 23 & 0.03 & {\bf 2.9}  &   EC \\
Cyg~A & HS-A & 0.55 & 4.0e4 & 0.8$\pm$0.2 & 19 & -- & 1.2
 & 180 & 1.0 & 1.3e2 & {\bf 0.98} & 5.0  &   SSC \\
      & HS-D & 0.50 & 5.0e4 & 0.8$\pm$0.2 & 29 & $<$8 & 1.2
 & 190 & 1.2 & 1.7e2 & {\bf 0.94} & 5.5  &   SSC \\
3C452 & L & 0.78 & 4.0e3 & 0.7$\pm$0.3 & 41 & -- & 80
 & 2.4 & 4.7e2 & 1.3 & 0.09 & {\bf 1.1}  &   EC \\
\enddata
\tablecomments{$Observables$ -- $\alpha_{\rm R}$: 
radio spectral index at 5 GHz, 
$f_{\rm R}$: radio flux density at 5 GHz in mJy,  
$\alpha_{\rm X}$: X-ray spectral index at 1 keV, 
$f_{\rm X}$:  X-ray flux density at 1 keV in nJy, 
$f_{\rm O}$:  optical flux density at 5$\times$10$^{14}$ Hz in $\mu$Jy, 
and $\theta$: radial size of the emitting region in arcsec. Parameters with 
suffix $f$ is assumed to be a listed value.
$Model$ $results$ -- $B_{\rm eq}(1)$; the equipartition magnetic field 
for no beaming $\delta$ = 1, $R_{\rm SSC}(1)$; the ratio of 
observed X-ray flux density to that expected from SSC model for $\delta$
 = 1,   $R_{\rm EC}(1)$; the ratio of 
observed X-ray flux density to that expected from EC model for
 $\delta$ = 1,  $\delta_{\rm SSC}$; the Doppler
beaming factor required to hold equipartition i.e., 
$R_{\rm SSC}$($\delta_{\rm SSC}$) $\simeq$ 1, and  $\delta_{\rm EC}$; 
the Doppler beaming factor required to hold equipartition for EC model, 
i.e., $R_{\rm EC}$($\delta_{\rm EC}$) $\simeq$ 1. class; Most likely
scenario of producing observed X-rays. 
}
\end{deluxetable}

\end{document}